# Combining BART and Principal Stratification to estimate the effect of intermediate variables on primary outcomes with application to estimating the effect of family planning on employment in Nigeria and Senegal.

Lucas Godoy Garraza[1][1], Ilene Speizer[2,3], and Leontine Alkema[1]

[1] Department of Biostatistics and Epidemiology, University of Massachusetts Amherst

[2] Department of Maternal and Child Health, University of North Carolina at Chapel Hill

[3] Carolina Population Center, University of North Carolina at Chapel Hill

[1] Contact: lgodoygarraz@umass.edu.

**Abstract**

There is interest in learning about the causal effects of modern contraceptive use on empowerment outcomes. Data on this question often come from family planning (FP) programs that increase access to FP and facilitate contraceptive use among some women, rather than directly assigning use. Women whose contraceptive behavior changes because of these programs ("compliers") may differ from target populations in ways that alter the consequences of contraceptive use for empowerment outcomes.

We propose a two-step approach. First, we use principal stratification and Bayesian Additive Regression Trees (BART) to estimate the effect of modern contraceptive use among compliers in the study population, treating the FP program as an instrument rather than as the treatment of interest. Second, we generalize these complier-specific effects to a broader population by averaging conditional effects over the covariate distribution in the target population, with uncertainty in that distribution quantified via a Bayesian bootstrap applied to external complex survey data.

We examine performance in simulation designs previously used to evaluate IV estimators. We then apply the approach to employment among urban women in Nigeria and Senegal, finding strong and heterogeneous effects of contraceptive use. Sensitivity analyses suggest robustness to violations of assumptions for internal and external validity.

## 1. Introduction

There is interest in learning the causal effect of family planning (FP) on empowerment-related outcomes such as employment. Prior research has suggested that as more women are able to control the timing, spacing, and number of children they have through modern family planning use, these same women are better able to attain higher education levels and engage more fully in the job market (Finlay, 2021; Joshi & Schultz, 2007). Examining this relationship with observational data is difficult because there are systematic differences between women who use modern contraceptives and those who do not, and these differences may well relate with empowerment.

While the use of FP cannot be randomized, there are studies randomizing "encouragement" (e.g., the provision of information or behavior change programming) or studies where such randomization might be thought to have occurred (at least approximately) after considering all observed covariates. In such setting, instrumental variable approaches and, in particular, principal stratification (Imbens & Rubin, 1997), can be used to identify the causal effect of primary interest—the effect of modern contraceptive use—at least for a certain segment of women, i.e., women for whom the encouragement induces a change in FP behavior (complier). Throughout, we treat the FP program as an encouragement (instrument) and the adoption of modern contraception as the treatment whose effects we estimate.

The segment of women among whom the effect can be identified, however, may differ systematically from the population of interest, and those differences may modify the effect of adopting modern contraception. The question of how to generalize results from a selective sample to a broader population of interest while accounting for differences in the distribution of effect modifiers has received some attention in the causal inference literature (Wang & Tchetgen

Tchetgen, 2018; Rudolph & Laan, 2017; Aronow & Carnegie, 2013; Angrist & Fernandez-Val, 2010; Hernán & Robins, 2006).

In this work we propose a two-step approach. In the first step, we use principal stratification in combination with Bayesian nonparametric regression to flexibly estimate the conditional average treatment effects (CATEs) in a selective sample (the source), where "treatment" refers to modern contraceptive use. We refer to the approach as Prince BART (Chen et al., 2024; Godoy Garraza et al., 2024). In the second step, we use a Bayesian bootstrap to estimate the distribution of covariates in a target population based on a complex probabilistic sample (the target). To obtain the population average treatment effect (PATE), we average the CATEs over the estimated distribution of covariates in the target population, as in the plug-in g-formula (Robins, 1986). Because this approach only accounts for shifts in observed covariates, we use sensitivity analysis to assess the possible impact of unobserved shifts. We refer to the entire strategy as Prince BART Generalized (PBG).

Conceptually and technically, our work is best viewed as an integrated application of existing ideas rather than a new estimation paradigm. Bayesian mixture models for principal strata and BART are both well established, and closely related BART-based principal-stratification models have recently been proposed for truncation-by-death settings (Chen et al., 2024). Our contribution is to tailor and combine these tools for instrumental-variables problems in which an intermediate variable affects a subsequent outcome, and where interest lies both in the complier (or principal-stratum) effect and its generalization to a broader target population. In particular, we focus on the empirically common case in which the instrument is only conditionally independent of the potential outcomes given covariates X, so that it can be regarded as randomized only within levels of X. In this setting, conventional instrumental-variables estimators that rely on linear models with additive covariate adjustment may be sensitive to functional-form assumptions and can perform poorly when relationships are nonlinear or effects are heterogeneous. By embedding a Bayesian principal-stratification mixture model within BART, and combining it with a Bayesian bootstrap and external data on the distribution of baseline covariates, we provide a flexible framework tailored to this niche: it accommodates complex dependence on observed characteristics, yields interpretable complier-type effects, and provides a framework for assessing how these effects may generalize to a target population .

Our paper is organized as follows. The next section introduces case studies that motivate the work followed by a section that reviews the relevant literature on instrumental variables/principal stratification estimation and generalization. The methods section introduces the approach to estimate CATEs in a selective sample and to obtain the PATEs using complex survey data from the target population. This section identifies assumptions needed for internal and external validity and provides approaches to gauge the sensitivity of the results to departures from those assumptions. We apply the approach to estimate the effect of modern contraceptive use on employment among broad populations of urban women in Nigeria and Senegal. Some simulations comparing Prince BART with alternative strategies are included in the Supplemental Materials. We end with a discussion of our approach and its limitations.

## 2. Case studies and data

The proposed methods are motivated by interest in the causal effect of FP on empowerment-related outcomes such as employment. Data on this effect is available from the Measurement, Learning & Evaluation (MLE) project (Carolina Population Center at the University of North Carolina in Chapel Hill, 2022). To generalize to broader populations of urban women in both countries, we use household survey data.

The MLE project examined the impact of the Urban Reproductive Health Initiative demand and supply-side interventions on FP outcomes in Kenya, Nigeria, Senegal, and the state of Uttar Pradesh, India. In both Nigeria and Senegal, the program was initially introduced in 2010/2011 in four cities (Abuja, Ibadan, Ilorin, and Kaduna in Nigeria and Dakar, Guédiawaye, Pikine and Mbao in Senegal) and after two years of implementation, the most effective strategies were adopted in two "delayed intervention" cities (Benin City and Zaria in Nigeria and Kaolack and Mbour in Senegal). Longitudinal data were collected prior to the start of the FP interventions (baseline), and four years later (endline). For the present work we focus on women participating in both administrations of the survey who, at baseline, had never used modern contraception (6,808 in Nigeria and 4,380 in Senegal). The outcome of interest is employment at endline, i.e., whether a woman was employed in the year prior to the endline survey. For details on the Nigeria and Senegal impact evaluation data see Atagame et al., (2017) and Benson et al. (2018).

In both countries, the set of study cities was determined by the implementing programs rather than by the evaluation team. Program staff report that they prioritized large urban areas with substantial populations and existing health infrastructure: in Nigeria, they selected large cities (excluding Lagos) across five states (with two cities in the same state), and in Senegal the four initial intervention sites are all within the Dakar region, while the two delayed sites lie outside Dakar. The staggered roll-out in Senegal appears to have been driven mainly by logistical convenience and concerns about contamination between nearby sites. These features suggest that assignment operated primarily at the programmatic and logistical level, but they also leave open the possibility of residual city-level confounding if unobserved city characteristics are not fully captured by our covariates. We therefore treat instrument validity as an assumption and assess robustness to such confounding using the sensitivity analysis described in Section 7.1.

Information on the covariate distributions among women residing in urban Nigeria or Senegal who have not used modern contraception and wish to avoid or delay pregnancy comes from the Demographic and Health Surveys (DHS, Corsi et al., 2012). The DHS is a standardized household survey conducted periodically on over 90 countries since 1984 with the support of the United States Agency for International Development (USAID). DHS is widely regarded as a reference source for population descriptive information in family planning and other health topics in low- and middle-income countries. For that reason, most variables used in MLE are also measured in the DHS and were operationalized in the same way.

We use data from the Nigeria and Senegal DHS carried out in 2018 and 2023, respectively, (National Population Commission - NPC & ICF, 2019; Agence Nationale de la Statistique et de la Démographie (ANSD) & ICF, 2024). The DHS are based on a complex sampling design. Specifically, a stratified two-stage probabilistic sample of approximately 42,000 and 8,800 households, in the case of Nigeria and Senegal respectively. The strata are defined by dividing each subnational jurisdiction (36 states in Nigeria, 14 regions in Senegal) into rural and urban areas. Independently for each stratum, a sample of compact geographic units referred to as primary sample units (PSUs) is selected with probability proportional to size (PPS), where information on the number of households is used as measure of size (1,400 in PSUs in Nigeria, and 400 in Senegal). After listing all households across the sampled PSUs, a systematic random sample of a fixed number of households per PSU is taken. DHS attempts to interview all women aged 15 to 49 residing in the selected households.  The publicly released dataset includes sample weights computed to reflect the inclusion and response probabilities; nonresponse adjustment varies only by strata (i.e., it does not involve poststratification or calibration).

## 3. Related literature

In this work we address two interrelated questions. The first question is how to estimate the effect of a "treatment" whose assignment may be considered unconfounded only in a very selective sample, and for which causal effects may vary across units. The second question is how to generalize the results from that selective sample to a broader population of interest. These aspects, namely internal and external validity, and how they relate to effect heterogeneity, have received considerable attention in the causal inference literature.

### *3.1 Estimating the complier effect*

Our approach to the first question is rooted in the modern potential-outcomes formulation of instrumental variables (IV) and in principal stratification (PS). In this formulation, introduced by Imbens & Angrist (1994) and Angrist et al., (1996), a binary instrument $Z$ shifts the probability of receiving an intermediate treatment $W$ and identifies the effect of $W$among compliers—individuals whose treatment status changes when the instrument changes. Principal stratification (Frangakis & Rubin, 2002) generalize this idea by treating compliance types as latent strata and defining complier (or "principal") causal effects as averages within the complier stratum. Within the broader principal-stratification literature, some methods rely instead on principal ignorability assumptions (Jo & Stuart, 2009). Here we explicitly remain in the IV framework and use the encouragement variable $Z$ for identification. Throughout, we work within this potential-outcomes/PS view of IV, rather than the older simultaneous-equations tradition.

Two-stage least squares (2SLS) remains the most widely used estimator of the complier effect in practice (Blandhol et al., 2022; Słoczyński et al., 2025). With a binary instrument and no covariates, 2SLS reduces to a simple ratio of differences and identifies an average complier effect under standard assumptions. When covariates $X$ are included linearly—a common practice when the instrument is assumed independent of potential outcomes only conditional on $X$—the resulting estimand generally combines effects across compliance types and across values of $X$ with weights that can be difficult to interpret and need not correspond to a clearly defined complier-specific parameter (Blandhol et al., 2022). Allowing more flexible specifications does not, in general, restore a clear complier interpretation (Słoczyński, 2024). In addition, the ratio form of the estimator is not intrinsically bounded by the outcome scale; in finite samples, small instrument-induced treatment differences can yield effect estimates that fall outside the logical range of a probability difference, an issue noted for other ratio-type estimators (e.g., Robins et al., 2007).

A broad class of alternatives to 2SLS builds on the instrument propensity score —the conditional probability of treatment assignment given covariates— and the identification results of Abadie, (2003). Abadie showed that complier-specific moments can be represented using κ-weights derived from the instrument propensity score, providing a flexible route to estimating complier means when the instrument is independent of potential outcomes conditional on $X$. Subsequent work has examined normalized and unnormalized κ-weighting estimators, their finite-sample behavior, and their sensitivity to outcome scaling and limited treatment uptake differences (Heiler, 2022; Słoczyński et al., 2025). These methods directly target complier-specific averages and allow flexible modeling of the instrument propensity, but by default they still yield a single average effect rather than a full characterization of heterogeneity, and, as ratio-based estimators, they share the lack of intrinsic boundedness.

A different line of work—closest to our approach—combines principal stratification with Bayesian mixture models. Imbens & Rubin (1997) introduced a fully Bayesian principal-stratification estimator and showed that modeling latent strata and potential outcomes jointly can improve finite-sample performance, partly by avoiding unstable ratio representations and enforcing logical constraints on probabilities. Subsequent contributions have extended this framework to various outcome types and settings (Hirano et al., 2000; Frumento et al., 2012; Liu et al., 2024), with user-friendly implementations such as the PStrata package (Liu & Li, 2023).

More recently, researchers have replaced parametric components in these mixture models with flexible nonparametric regression. A particularly attractive alternative is Bayesian Additive Regression Trees (BART, Chipman et al., 2007, 2010). Chen et al. (2024) use this approach to estimate heterogeneous survivor causal effects in a critical care trial with truncation by death, using BART to flexibly model both latent stratum membership and potential outcomes. See also Kim & Zigler (2025). BART and related tree-based ensembles have become popular in causal inference because they capture complex nonlinearities and interactions while providing regularized estimates of conditional treatment effects (Hill, 2011; Hill et al., 2020). Compared to 2SLS, κ-weighting, and parametric Bayesian PS models, PS + BART—what we term Prince BART—is particularly appealing in settings with conditionally independent instruments and substantial effect heterogeneity: it allows flexible adjustment for high-dimensional $X$, avoids some of the finite-sample instabilities associated with ratio-type estimators, and yields a full posterior distribution of complier effects as a function of covariates, which we leverage in the generalization step.

### *3.2 Generalizing from a selective sample*

This paper also concerns generalizing from a study sample to a target population. Prior literature has considered how to generalize from studies that use principal stratification or instrumental variables to identify the causal effect of interest, including trials with imperfect compliance or randomized encouragement designs (Angrist & Fernandez-Val, 2010; Aronow & Carnegie, 2013; Hernán & Robins, 2006; Rudolph & Laan, 2017; Wang & Tchetgen Tchetgen, 2018). Collectively, this work has clarified the assumptions required for generalization from such studies. There is also a growing body of literature on generalizing or transporting results from randomized trials that consider the issue that participants are rarely a probabilistic sample from the population of interest (recent reviews include Degtiar & Rose (2023) and Colnet et al. (2023); the topic is also discussed in Li et al., (2023) from a Bayesian perspective). Both Angrist & Fernandez-Val (2010) and Rudolph & Laan (2017) consider ways to address observed covariate shifts across populations.

Our setting and approach are somewhat different from existing work in two ways. Firstly, we only assume access to covariate information from the target population. In particular, we do not assume any information on the instrument in the population. Secondly, we aim to leverage flexibly estimated CATEs from the source sample, i.e., CATEs that were estimated from combining principal stratification with BART. In our approach, we combine CATEs with external information on the distribution of covariates in the target population, using a Bayesian bootstrap to model this distribution flexibly. An early proposal to combine CATEs estimated with BART with covariate information from the target population is discussed in Appendix A of Hill (2011). Hill proposed to either ignore the additional estimation uncertainty in the covariate density (if the external sample was very large) or use a frequentist bootstrap to incorporate it. Our approach builds off several recent applications that have used the Bayesian counterpart, i.e., the Bayesian bootstrap (BB), to incorporate the uncertainty introduced by the estimation of the covariate distribution in the target population while avoiding the need to specify a full parametric model (e.g., Oganisian et al., 2022; Taddy et al., 2016; Wang et al., 2015; Xu et al., 2018).

We use a modified version of the Bayesian bootstrap to account for the complex sampling design associated with the covariate data in the target population. In its original formulation, the BB as introduced by Rubin (1981) (see also Chamberlain & Imbens, 2003, for an early application) assumes the data is independent and identically distributed (i.i.d.) generated. This approach is not appropriate in our application, due to the complex sampling design that was used to collect the covariate data on the target population. How to address complex sampling design in Bayesian estimation has received considerable attention (see Little, 2014, for an interesting account). Some extensions of the BB to handle complex samples have been proposed, including sampling from finite population (Lo, 1988), unequal probability sampling (Cohen, 1997; Rao & Wu, 2010; Zangeneh & Little, 2015), and the combination of stratification and clustering (Aitkin, 2008; Dong et al., 2014; Makela et al., 2018). To the best of our knowledge, these innovations have not yet been applied to causal inference.

## 4. Methods

Our analysis proceeds in two main stages, summarized in Figure 1. In the first stage, we focus on internal validity using data from the MLE source study. We treat assignment to the FP program as an instrument and use a Bayesian principal–stratification model, implemented with BART (Prince BART), to (i) estimate for each woman the probability of being a complier—someone who adopts modern contraception because of the program—and (ii) model the employment outcome as a function of baseline covariates, the latent principal stratum, and assignment. This yields posterior draws of complier-specific conditional average treatment effects, $CATE_C(x)$, within the source sample. In the second stage, we address external validity by combining these $CATE_C(x)$ estimates with covariate data from a large, complex survey of the target population (DHS). We use a scaled Bayesian bootstrap over PSUs and survey weights to model the distribution of covariates in the target population and average the predicted $CATE_C(x)$ accordingly, obtaining posterior draws of the population average treatment effect (PATE). The sections that follow describe the setup, identification assumptions, mixture model, BART specification, and scaled Bayesian bootstrap in more detail.

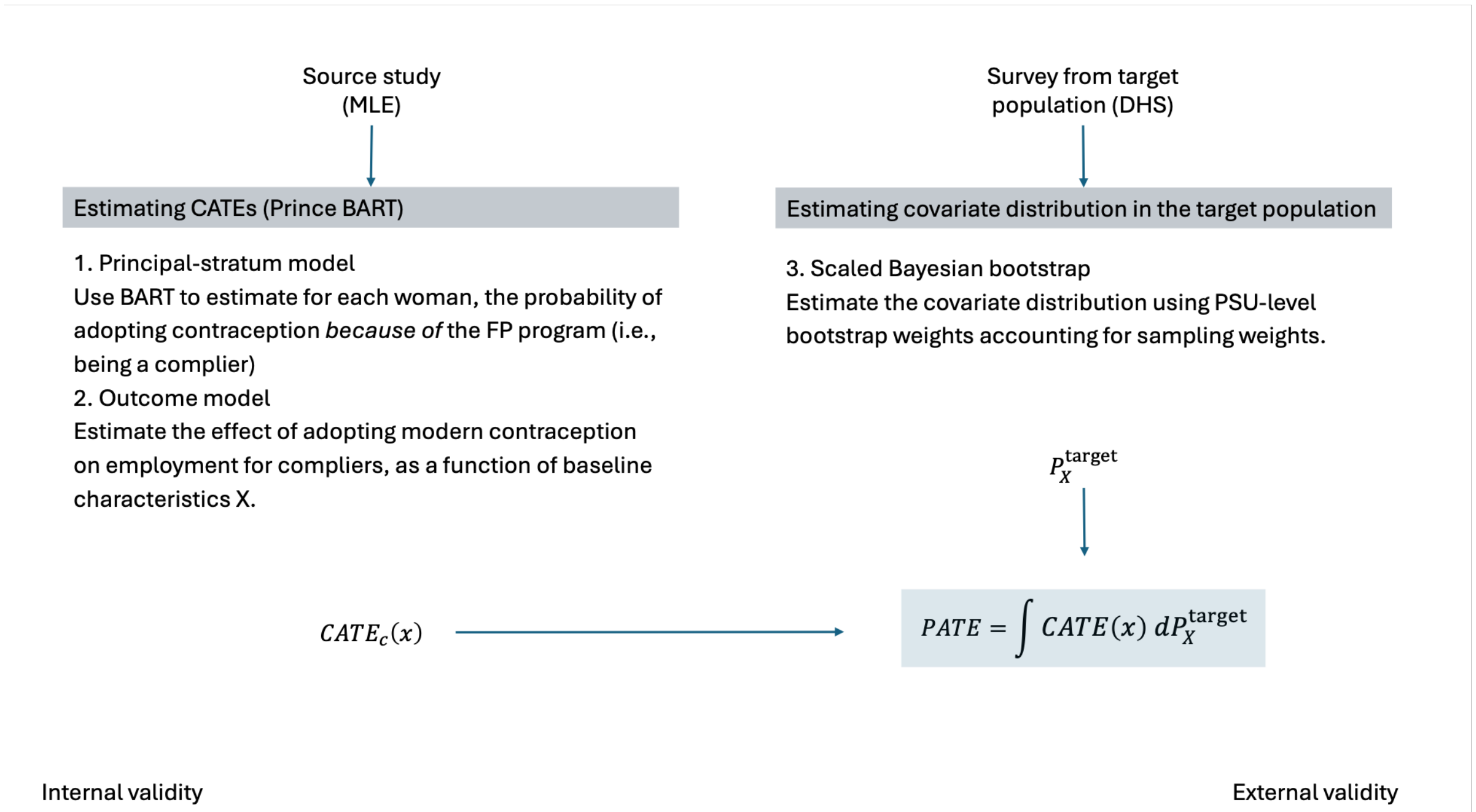

***Figure 1. Overview of the Prince BART Generalized procedure****. The left panel (internal validity) shows how data from the MLE source study are analyzed using Prince BART. A BART principal-stratum model is used to estimate, for each woman, the probability of adopting modern contraception because of the FP program (being a complier), and a BART outcome model is used to estimate the effect of adoption on employment as a function of baseline covariates X, yielding posterior draws of complier $CATE_C(x)$. The right panel (external validity) shows how these $CATE_C(x)$ estimates are combined with covariate data from a large DHS survey of the target population. A scaled Bayesian bootstrap over PSUs and survey weights models the covariate distribution $P_X$ in the target population and averages the predicted effects to obtain posterior draws of the population average treatment effect (PATE).*

### *4.1 Set up and notation*

For a sample of units indexed by $i = 1, \dots, n$, termed the source sample, $Z_i = \{0,1\}$ indicates assignment to some intervention or program encouraging a FP behavior; $W_i = \{0,1\}$, indicates whether the subsequent adoption of the FP behavior occurs; and $Y_i = \{0,1\}$ is the outcome of primary interest. In our application, the source sample comes from the MLE program and is composed by women in 6 cities in Nigeria and 6 in Senegal who had never used modern contraception at baseline, $Z_i$ indicates whether the woman resided in one of 4 cities exposed to a FP program two years earlier (intervention cities) rather than the two cities adopting the program two years later (control cities); $W_i$ indicates adoption of modern contraception after baseline, $Y_i$ indicates employment in the 12 months before endline. We denote by $X_i$ a set of measured baseline characteristics (e.g., religion, marital status, age, education, wealth, parity) including baseline values of the outcomes (work the year before baseline). We use $U_i$ to refer to an unobserved covariate which may relate to both the $W_i$ and $Y_i$ (e.g., motivation to work).

Depending on assignment, $Z_i$, and the subsequent FP behavior, $W_i$ , there are 2 potential values for $W_i^*(Z_i)$, and 4 potential outcomes, $Y_i^*\big(Z_i, W_i^*(Z_i)\big)$, of which we can only possibly observe the ones corresponding to the actual assignment, i.e., $W_i = W_i^*(\mathrm{z})$ and $Y_i = Y_i^*\big(\mathrm{z}, W_i^*(\mathrm{z})\big)$ for $z = 0, 1$. Note that this notation does not accommodate dependence of the potential outcomes on the values for other units or hidden versions of the treatment (i.e., the stable unit treatment values assumption, Rubin, 1974).

#### *4.1.1 Principal stratification*

Based on the potential values $W_i^*(Z_i)$, we can define the following latent partitions in the sample,

$$G_i^* = \mathrm{g}\big(W_i^*(0), W_i^*(1)\big) = \begin{cases} Never-takers\,(n), & if\ W_i^*(z) = 0, \\ Compliers(c), & if\ W_i^*(\mathrm{z}) = \mathrm{z}, \\ Defiers(d), & if\ W_i^*(\mathrm{z}) = 1-\mathrm{z}, \\ Always-takers(a), & if\ W_i^*(\mathrm{z}) = 1, \end{cases} \quad (1)$$

for z = 0,1. The latent class $G_i^*$ is a covariate, i.e., a pre-treatment variable. Unlike other covariates, however, $G_i^*$ is only partially observed: if we cross-tabulated each woman based on the observed values of $w$ and z, each cell encompasses a mixture of two strata.

#### *4.1.2 Target population*

For the application, the population of interest are women residing in urban Nigeria or Senegal who have not used modern contraception and wish to avoid or delay pregnancy. Continuing with $X_i$, $W_i$ , and $Y_i$ , as defined above, we let $i = n + 1, \dots, n + m$ index the units (here the women) in a population of interest, and we let $T_i$ indicate if a woman was part of the source sample or is part of the target population, with $T_i = 1$ if $i > n$ and zero otherwise. For a probabilistic sample of women from the target population, we observe a set of covariate characteristics, $\ddot{X}_i \subseteq X_i$, containing most of the same set of covariate characteristic observed in the source sample. The values for the variables that are only observed in the source sample, i.e., $X_i \setminus \ddot{X}_i$, are imputed as explained in the section on generalizing to the target population.

### ***4.2 Estimand***

Define $Y_i^*(w)$ for $w = 0{,}1$ as the potential outcome if a woman were to (or were not to) adopt modern contraceptives and the *individual treatment effect* (ITE) as the contrast between these two potential outcomes. We are interested in the average ITE among a target population, i.e., the *population average treatment effect* (PATE) defined as

$$\text{PATE} \equiv \mathbb{E}(Y_i^*(1) - Y_i^*(0) | T_i = 1), \qquad (2)$$

where the expectation is taken over the entire target population of interest. In other words, the average effect if everyone in the population of interest were to adopt modern contraception as opposed to no one. An equivalent expression for the PATE is,

$$\text{PATE} = \mathbb{E}_{X|T=1}(\text{CATE(x)}) = \int \text{CATE}(x) dP_{X|T=1}(x), \tag{3}$$

where $CATE(x) \equiv \mathbb{E}(Y_i^*(1) - Y_i^*(0)|X_i = x)$ is the conditional average treatment effect and $P_{X|T=1}(x)$ the distribution of covariates in the population of interest.

While our ultimate interest lies on the PATE, we will initially focus on estimating the $CATE$'s among women affected by a FP program, i.e., the *conditional average treatment effects among compliers* ($CATE_c$'s), defined as

$$CATE_c(x) \equiv \mathbb{E}(Y_i^*(1) - Y_i^*(0)|X_i = x, G_i^* = c). \tag{4}$$

We assume, as further discussed below, this is the only segment providing information on the effect of interest.

### *4.3 Identifying assumptions*

Box 1 includes the assumptions needed to identify the causal effect in a sample of woman affected by a FP program (internal validity) and assumptions required to generalize or transport the conclusion to a broader population of interest (external validity).

INTERNAL VALIDITY

***Assumption 1 (Conditional unconfoundedness).***
$P\left(Z_i|X_i, W_i^*(1), W_i^*(0), Y_i^*(0{,}0), Y_i^*(1{,}1), Y_i^*(0{,}1), Y_i^*(1{,}0)\right) = P(Z_i|X_i)$, i.e., $Z_i$ was assigned independently of the potential outcomes after considering baseline differences on observed covariates, $X_i$.

***Assumption 2 (Overlap).***

$0 < P(Z_i = 1|X_i) < 1$, or just $P(Z_i = 1|X_i) < 1$, if we focus on the effect among the treated, i.e., the assignment is not a deterministic function of the baseline covariates.

***Assumption 3 (Monotonicity).***

$W_i^*(1) \geq W_i^*(0)$, i.e., there are no defiers.

***Assumption 4 (Exclusion restriction).***

$Y_i^*(0, w) = Y_i^*(1, w)$ for $w = 0, 1$, i.e., there is no direct effect of Z on the outcome.

EXTERNAL VALIDITY

***Assumption 5 (Conditional transportability).***

$\mathbb{E}(Y_i^*(1) - Y_i^*(0)|X_i, G_i^* = c) = \mathbb{E}(Y_i^*(1) - Y_i^*(0)|X_i)$, i.e., there are no unmeasured effect modifiers. Assumption 5 holds if either assumption 5a or 5b is met:

*Assumption 5.a* $P(G_i^* = c|\ X_i, U_i) = P(G_i^* = c|\ X_i)$, i.e., the unobserved confounder does not predict compliance conditionally on observed covariates.

*Assumption 5.b* $\mathbb{E}(Y_i^*(1) - Y_i^*(0)\ |X_i, U_i) = \mathbb{E}(Y_i^*(1) - Y_i^*(0)|X_i)$, the unobserved confounder does not modify the effect conditionally on observed covariates.

***Assumption 6 (Included support).***

$supp\left(P_{X|T=1}(X)\right) \subseteq supp\left(P_{X|\ C^*=1}(X)\right)$, i.e., the combination of covariate values in the target population is also present among compliers in the source sample.

*Box 1: Identifying assumptions.*

### *4.3.1 Internal validity*

The assumptions necessary for identification in this setting were first laid out in Angrist et al., (1996), who showed that, within a subpopulation of women with the same characteristics $\{i: X_i = x\}$, the assumptions ensure identification of complier-specific effects. The first two assumptions—*conditional unconfoundedness and overlap*—are standard in both instrumental-variables analyses and settings based on selection on observables.

In our context, assignment, Z, (in contrast to the actual treatment, W) was independent of individual-level motivation or family planning preferences. Nevertheless, clustered assignment with a small number of cities raises concerns about residual city-level confounding. We therefore rely on a rich set of covariates X, intended to capture the main predictors of employment Y and contraceptive use W—including baseline values of these outcomes. We examine how sensitive our findings are to violations of conditional unconfoundedness arising from unobserved city-level factors (Section Robustness Checks).

While overlap cannot be verified exactly, we can enforce practical overlap by restricting attention to covariate values for which the estimated probabilities $\Pr(Z_i = |X_i)$ are not extreme. In the empirical analysis, we estimate these probabilities and trim observations with very low or very high values, as described in Section 4.4.3.

Assumptions related to *monotonicity* and the *exclusion restriction* cannot be tested directly, but they appear substantively plausible in this setting: we do not expect increased access to family planning to reduce adoption of modern contraception, nor to affect employment prospects except through changes in family-planning behavior.

### *4.3.2 External validity*

Conditional transportability is the strongest assumption needed to generalize the results. This assumption is discussed in by Angrist & Fernandez-Val (2010), under the rubric “conditional effect ignorability”, and Aronow & Carnegie (2013), who termed it “latent ignorability of compliance with respect to treatment effect heterogeneity”. We note that the assumption is formally similar to one used to generalize or transport results from an RCT with perfect compliance, with $G_i^* = c$ set to $(1 - \mathrm{T}_i)$ (Colnet et al., 2022, 2023).

As discussed by Wang & Tchetgen Tchetgen (2018),[2] one of two conditions suffice for assumption 5 to hold. The first condition (5a) states that, among units with the same covariate values, the compliers (or the study participants, when the source study is an RCT with perfect compliance) are a random sample of the population. The second condition (5b) states that the observed covariates are the only source of effect heterogeneity. Unlike the first one, the second condition does not rule out “selection bias”, i.e., difference between compliers and target population with respect to unobserved confounders. It does rule out, however, “gain-driven selection”, i.e., selection associated with the anticipated effect size (Angrist & Fernandez-Val, 2010). Different versions of the assumption 5b, such as limiting effect homogeneity to the subsample of women with $W_i = w$, are considered by Hernán & Robins (2006)- termed “no current treatment interaction”.

In our application, assumption 5 requires that women affected by longer exposure to a FP program (i.e., compliers), differ from other women only with respect to covariates observed at baseline, unobserved characteristics that do not affect the outcomes, or unobserved characteristics that affect their probability of being employed at endline in the same way, irrespective of whether they adopt modern contraception.

[2] Wang & Tchetgen Tchetgen (2018) use a more general formulation of this assumption that allows no causal instrumental variables.

Assumption 6 (Included support) is related to assumption 2. We can always limit the source sample to "enforce" assumption 2 but this will affect the validity of assumption 6. We introduce sensitivity checks in a later section to examine the consequences of departures from these assumptions.

### *4.4 Estimating the effect among women affected by the FP program*

While assumptions 1 through 4 suffice to ensure nonparametric identification of the $CATE_c$, with many covariates (and/or covariates with many values), it is convenient to introduce statistical models. We adopted a non-parametric version of the mixture-model based Bayesian approach first introduced by Imbens & Rubin (1997) for the analysis of randomized trials with noncompliance.

#### *4.4.1 Overview of mixture model and estimation process*

For each unit in the source sample, we observed 4 random variables $\{Y_i, Z_i, W_i, X_i\}$. We assume the joint distribution of these variables is governed by a generic parameter $\theta$, with prior distribution $p(\theta)$, conditional on which the random variables for each unit are i.i.d. Let $\mathcal{G}(z, w)$ denote the set of principal strata compatible with each combination of $(Z, W)$, e.g., $\mathcal{G}(1,1) = \{c, a\}$. Then the likelihood of the observed data is proportional to (see Appendix I for additional details):

$$\prod_{i=1}^{n} P(X_i, Z_i, W_i, Y_i|\theta) \propto \prod_{i=1}^{n} \sum_{g \in \mathcal{G}(Z_i, W_i)} P(G_i^* = g|\, X_i, \theta_G) P(Y_i|G_i^* = g, Z_i, X_i, \theta_Y)\,. \tag{5}$$

This suggests we need to specify two models: (i) a principal strata model, denoted by $\pi_g(x) \equiv P(G_i^* = g|X_i, \theta_G)$, and (ii) an outcome model, denoted by $\varpi_{gz}(x) \equiv P(Y_i|G_i^* = g, Z_i, X_i, \theta_Y)$. For Bayesian inference, we further need to specify prior distribution of the parameters governing these models. Theses specifications will be discussed in the next section.

Given the models and prior for the model parameters, we can approximate the posterior distribution of the causal estimands (i.e., quantities that depend on $\pi$'s and the $\varpi$'s), despite the fact that $G^*$ is missing for the subset units $\{i: Z_i \neq W_i\}$. We use a data augmentation (DA) approach; the algorithm is described in Appendix I.

We have yet to specify the models for $\pi_g(x)$ and $\varpi_{zg}(x)$. The most common choice is to use generalized linear models. Specifically, given the discrete nature of the latent class and the outcome, logistic regression is a common specification. In this article, we use a much more flexible option, Bayesian Additive Regression Trees (BART, Chipman et al., 2007, 2010). This alternative was recently proposed by Chen *et al.* (2024) in a related context.

#### *4.4.2 Bayesian Additive Regression Trees*

We use a Bayesian nonparametric regression model based on an ensemble of trees, BART, to model class membership and the outcome given class membership as a function of covariates. Specifically, we model the probability of being a complier, the probability of being an always-taker rather than a never-taker among noncompliers, and the outcome probability conditional on $G^*$ and $Z$ as functions of covariates, each following a (probit) BART, i.e.,

$$bart^{(\ell)}(x) = \Phi\left(\sum_{j=1}^{J} h\left(x; T_j^{(\ell)}, M_j^{(\ell)}\right)\right), \quad (6)$$

for $\ell \in (\pi_c, \pi_a/(\pi_a + \pi_n), \varpi_{1c}, \varpi_{0c}, \varpi_{1a}, \varpi_{0n})$. Here $T_j^{(\ell)}$ is a set of rules splitting the covariate space into non-overlapping regions ("leaves"), $M_j^{(\ell)}$ is the corresponding set of leaf parameters, and h(.) is a step function linking both. A prediction is obtained by summing over many trees (e.g., $J = 200$).

To avoid overfitting, BART places a regularization prior on $\left(T_j^{(\ell)}, M_j^{(\ell)}\right)$ such that each tree contributes only a small part to the overall fit. The prior on the splitting rules, $p(T_j^{(\ell)})$, makes large, deep trees very unlikely, while the prior on the leaf parameters, $p\left(M_j^{(\ell)} \middle| T_j^{(\ell)}\right)$, shrink them towards a common value. Posterior samples of $\left\{T_j^{(\ell)}, M_j^{(\ell)}\right\}_{j=1}^{J}$ are obtained using Bayesian backfitting.

To minimize the risk of BART regularization-induced confounding (Hahn et al., 2020), we augment the covariates in these models with an estimate of the instrument propensity. We use a BART model to estimate this propensity score, i.e., $\hat{e}_i = \hat{P}(Z_i = 1|X_i)$. We obtain this estimate from a separate probit-BART model for the instrument,

$$\text{e(x)} \equiv P(Z_i = 1|X_i = x) = bart^{(e)}(x). \quad (7)$$

Thus, $\hat{e}_i = bart^{(e)}(X_i)$ enters as an additional predictor in the principal-stratum and outcome models. Additional details on the BART specification, priors, and fitting algorithms are provided in Chipman et al. (2007, 2010), Hill et al. (2020) and in Appendix II.

#### 4.4.3 *Enforcing practical overlap*

As discussed in the internal validity section, exact overlap cannot be verified but we can enforce practical overlap by excluding observations with extreme assignment probabilities. We use the instrument model $bart^{(e)}(x)$ described above to estimate $\hat{e}_i$ and restrict the analysis to the sample to women with $\hat{e}_i \in [.1, .9]$, following the rule of thumb of Crump et al. (2009). All subsequent analyses (estimation of the complier-specific effect and transport models) are conducted on this trimmed sample.

#### 4.4.4 *Summarizing effect heterogeneity*

While BART is more flexible than logistic regression, its outputs are also less easy to interpret. A general strategy to summarize complex "black box" models is to fit simpler, surrogate models (Molnar et al., 2020) to the predicted values of the complex model. We use a variation of this strategy, termed surrogate shallow tree, to identify combinations of predictors defining segments with relatively homogenous CATEs. Specifically, we fit a single parsimonious regression tree to BART predicted values. We ensure parsimoniousness by limiting the tree depth to 3. A similar procedure to examine effect heterogeneity is used in Logan et al., (2019), for example. Details are provided in the Appendix II, Section 6.

Once the segments are identified, we examine effect heterogeneity by computing the average CATE for each segment. Such an average is a "mixed" quantity, in the sense that it combines population parameters with the empirical distribution of covariates in the source (Li et al., 2023). Let $\mathcal{J} \equiv \{i: X_i^K = a\}$ denote a segment of the sample sharing one or a few baseline characteristics, say $X^K \subseteq X$. The *compliers CATE* is given by

$$MCATE_c(\mathcal{J}) \equiv \frac{1}{\sum_{i:i\in\mathcal{J}} \pi_c(x_i)} \sum_{i:i\in\mathcal{J}} CATE_c(x_i)\, \pi_c(x_i). \tag{8}$$

where $\pi_g(x) \equiv P(G_i^* = \mathrm{g}|X_i)$ is the probability of belonging to class g conditional on baseline characteristics. We can also average the $CATE_c$'s over the entire sample of compliers to obtain the *mixed average treatment effect among compliers* ($MATE_c$),

$$MATE_c \equiv \frac{1}{\sum_i \pi_c(x_i)} \sum_i CATE_c(x_i)\, \pi_c(x_i). \tag{9}$$

### *4.5 Generalizing results to a target population*

Given assumptions 5-6, expression (3) can be written as,

$$\mathrm{PATE} = \int \mathrm{CATE}(x) dP_{X|T=1}(x) = \int \mathrm{CATE_c}(x) dP_{X|T=1}(x). \tag{10}$$

Since the Bayesian mixture model results in estimates of the $\mathrm{CATE_c}$'s, estimating PATE reduces to estimating the distribution of covariates in the population of interest, $P_{X|T=1}(x)$. We prefer an approach that avoids posing a full parametric specification for the multidimensional X. In addition, the approach needs to take account of the covariate data being collected through a complex sampling design. We use an extension of Bayesian bootstrap (BB, Rubin, 1981) to accomplish these goals, referred as "scaled" BB.

The scaled BB is based on PSUs or clusters, which are the compact geographic areas at the first sampling stage. Let $q = 1, \dots, l$ index the sampled clusters and $j = 1, \dots, n_q$ the women sampled in each cluster. Let the variable $Q_j \in \{1, \dots, l\}$ indicate if the $j^{th}$ woman resides in $q$ cluster. The expression for the PATE using the scaled BB is obtained as follows:

$$PATE_{\{BB\}} = \sum_{q} \varphi_q \cdot f_q \cdot CATE_q^{PSU}. \quad (11)$$

where $\varphi_q$ , $\sum_q \varphi_q = 1$, refers to the bootstrap weight for cluster $q$, $f_q = n_q \cdot w_q$, is the weighted number of observations in the cluster, and $CATE_q^{PSU}$ refers to the average CATE in the $q$-th cluster, i.e., $CATE_q^{PSU} = \frac{1}{n_q} \sum_{j:Q_j=q} CATE(x_j)$. The Bayesian bootstrap is obtained by posing an improper Dirichlet proportional to $\prod_q \varphi_q^{-1}$ (sometimes termed Haldane prior). To obtain posterior draws, e.g., the $d^{th}$ draw $PATE_{BB}^{[d]}$, we combine the $d^{th}$ draw of bootstrap weight $\varphi_q^{[d]}$ with the $d^{th}$ posterior sample $CATE(x)^{[d]}$.

The approach depends on the weighted number of observations $f_q$. By introducing the $f_q$ we are implicitly using a pseudo likelihood proportional to $\prod_q \varphi_q^{n_q \cdot w_q}$ as in Rao & Wu (2010). The procedure can be also thought of as adjusting for the probability of sampling clusters of different sizes (Makela et al., 2018; Zangeneh & Little, 2015). A small simulation study comparing this BB approach with standard design-based estimation is included in Appendix VII. Both approaches show similar performance (in terms of frequentist operating characteristics).

#### *4.5.1 Missing covariates*

Four binary covariates capturing attitudes towards family planning were measured in the MLE source data but not collected in the DHS target data. These attitude indicators correspond to components of $X_i \setminus \ddot{X}_i$, the covariates observed in the source sample but missing in the target population. In our main analysis we treat these attitude indicators as latent for DHS women and integrate over them within the posterior simulation. For each attitude variable, $A_k$, we model

$$Pr\left(A_{ik} = 1 | \ddot{X}_i\right) = bart^{(A_k)}\left(\ddot{X}_i\right), \qquad (12)$$

where the BART model is fitted in the MLE data using all other covariates, $\ddot{X}_i$, as predictors. Posterior predictive draws from this model, $A_{ik}^{[d]}$, are used to construct the complete covariate vector $X_i^{[d]} = \left(\ddot{X}_i, A_{i1}^{[d]}, \ldots, A_{iK}^{[d]}\right)$ for each DHS respondent. The completed covariates $X_i^{[d]}$ are then used to evaluate $CATE^{[d]}\left(X_i^{[d]}\right)$.

This BART-based multiple-imputation step propagates both binomial uncertainty and model uncertainty about these attitudes into the posterior for the PATE. As a sensitivity analysis, we also consider an extreme "conservative" scenario in which all DHS women are assigned the attitude pattern associated with the smallest estimated complier effect, corresponding to the approach used in our original analysis.

## 5. Simulation study

In this section we use simulations to assess the finite-sample performance of Prince BART. To reduce researcher discretion, we adopt the data-generating processes (DGPs) devised by Heiler (2022) and recently reused by Słoczyński et al. (2025) in a similar context. The DGPs are detailed in Box 2. Our setup is identical to theirs, except that we dichotomize the outcome to better match our empirical application.

Design A corresponds to a fully independent instrument, while in the remaining designs the instrument is conditionally independent. In Designs A and B, treatment effect heterogeneity arises solely from the correlation between $\varepsilon_i^1$, the error term in the potential outcome of the treated, and $v_i$, the error term in the treatment equation. In Designs C and D, the dependence of $\mu_{y1}(X_i)$, the systematic component of the treated outcome, on X provides an additional source of heterogeneity through covariate-driven variation in treatment effects. Finally, in Design D, $\mu_z(X_i)$ , the systematic component of instrument assignment, is quadratic in X. Across all designs, the parameter $\delta \in \{0.01, 0.02, 0.05\}$ controls how close the instrument propensity score $e(X_i)$ can get to 0 or 1, and thus indexes the degree of covariate overlap between units with $Z_i = 0$ and $Z_i = 1$.

We compare Prince BART with three alternative estimators. The linear 2SLS estimator with X included additively ("2SLS") is the most common approach in applied work (e.g., Angrist & Pischke, 2009)). It serves as a benchmark and is expected to perform well in Designs A and B, but not necessarily elsewhere (Blandhol et al., 2022; Słoczyński et al., 2025). The normalized $\kappa$ weighting estimator ("$\kappa$-weighting (normalized)") was developed specifically to address shortcomings of 2SLS in settings like Designs C and D; we implement it using the covariate balancing propensity score, an approach shown to outperform other weighting estimators (Heiler, 2022, Słoczyński et al., 2025). Finally, a Bayesian principal stratification model ("PStrata") in which the treatment and outcome models are specified as logistic regressions, with X entering additively on the logit scale (Imbens & Rubin, 1997; Liu & Li, 2023). Compared with 2SLS, PStrata uses a link function that is more appropriate for the discrete variables involved and does not rely on a ratio estimator, but it remains relatively inflexible because all components are based on parametric logistic regression.

We generate the instrument, treatment, and potential outcomes using additive latent-index models with covariate-dependent systematic components and idiosyncratic shocks.

**Instrument**

$$Z_i = 1\big(u_i < e(X_i)\big), \qquad e(X_i) = 1/1 + \exp(-\mu_z(X_i)\theta_0)$$

where $\theta_0 = \log\left(\frac{1-\delta}{\delta}\right)$ and $\delta \in \{0.01, 0.02, 0.05\}$ controls covariate overlap.

**Treatment (latent index)**

$$W_i^*(z) = 1(\mu_w(X_i, z) > v_i),$$

where $v_i$ is an term in the treatment equation.

**Potential outcomes (latent indices)**

$$Y_i^*(1) = 1\big(\mu_{y1}(X_i) + \varepsilon_i^1 > \tilde{s}\big), \qquad Y_i^*(0) = 1(\varepsilon_i^0 > 0),$$

where $\mu_{y1}(X_i)$ is the systematic component of the potential outcome of the treated, $\varepsilon_i^0, \varepsilon_i^1$ are error terms, and $\tilde{s}$ is the sample median of $\mu_{y1}(X_i) + \varepsilon_i^1$ in each simulated dataset.

**Covariates and error terms**

$$X_i \sim \text{Unif}(0,1), \qquad u_i \sim \text{Unif}(0,1), \qquad (\varepsilon_i^0, \varepsilon_i^1, v_i)' \sim \text{N}(0, \Sigma), \qquad \Sigma = \begin{pmatrix} 1 & 0 & 0 \\ 0 & 1 & 0.5 \\ 0 & 0.5 & 1 \end{pmatrix}.$$

Correlation between $\varepsilon_i^1$ and $v_i$ induces heterogeneity across compliance types.

**Design-specific systematic components**

| Design | $\mu_w(x, z)$ | $\mu_{y1}(x)$ | $\mu_z(x)$ |
|---|---|---|---|
| A | $4z$ | $0.3989$ | $2x - 1$ |
| B | $-1 + 2x + 2.122z$ | $0.3989$ | $2x - 1$ |
| C | $-1 + 2x + 2.122z$ | $9(x + 3)^2$ | $2x - 1$ |
| D | $-1 + 2x + 2.122z$ | $9(x + 3)^2$ | $x + x^2 - 1$ |

We consider two sample sizes, $N \in \{500, 1000\}$, and 400 Monte Carlo replications for each combination of a design, a value of δ, and a sample size.

*Box 2: Data-generating process*

All simulations were implemented in R. Two-stage least squares was fit using the `ivreg` package (Fox et al., 2025), κ–weighting estimators using the `kappalate` package (Uysal et al., 2023), and Bayesian parametric principal–stratification models using the `PStrata` package. Prince BART was implemented in a publicly available R package (Godoy Garraza, 2025), which wraps the `darts` engine (Dorie et al., 2025) in a discrete sampler for latent strata. For the BART components we used default tree priors in Chipman et al. (2010), a probit link for binary outcomes, and 200 trees. For each design we ran 1,500 iterations, discarding the first 750 as warm-up iterations and retaining 750 posterior draws, with these choices guided by effective sample size and $\hat{R}$ diagnostics from pilot runs. The computation was fast for all estimators except Prince BART. The 2SLS, κ–weighting, parametric principal stratification (PS), and PStrata models ran in negligible time per dataset. Prince BART is more computationally intensive, particularly in the simulation study where it must be re-estimated across 400 replications. In the empirical applications, where models are fit once per country rather than hundreds of times, a full Prince BART analysis takes approximately 10–30 minutes on a standard laptop or desktop computer. Full code is publicly available (Godoy Garraza, 2025, https://github.com/AlkemaLab/prince_BART).

Figure 2 and Figure 3 summarize relative MSE and coverage, respectively (additional results are in Appendix VI). As expected, 2SLS outperforms $\kappa$ weighting in scenarios A and B: in these settings, the additional flexibility of $\kappa$ weighting is unnecessary and only increases variance. In Designs C and D, by contrast, $\kappa$ weighting outperforms 2SLS: the extra flexibility reduces bias sufficiently to more than offset the increase in variance. These findings are fully consistent with the simulation results in Słoczyński et al. (2025). PSstrata generally improves on 2SLS in terms of MSE, outperforming it in all but one d setting. In the most demanding settings (Designs C and D), PStrata lags behind $\kappa$ weighting. Across the four designs, Prince BART is either the best performer or very close to the best performer.

Turning to coverage (Figure 3), performance differences are more pronounced in Designs C and D. In the simpler Designs A and B, all estimators achieve coverage close to nominal levels. In contrast, in Designs C and D, 2SLS and PStrata exhibit coverage that is substantially below nominal levels, reflecting bias under model misspecification. Coverage for κ-weighting and Prince BART is broadly similar across designs, with both remaining close to nominal levels in most scenarios. Taken together with the MSE results, these findings indicate that Prince BART achieves lower or comparable MSE without sacrificing coverage.

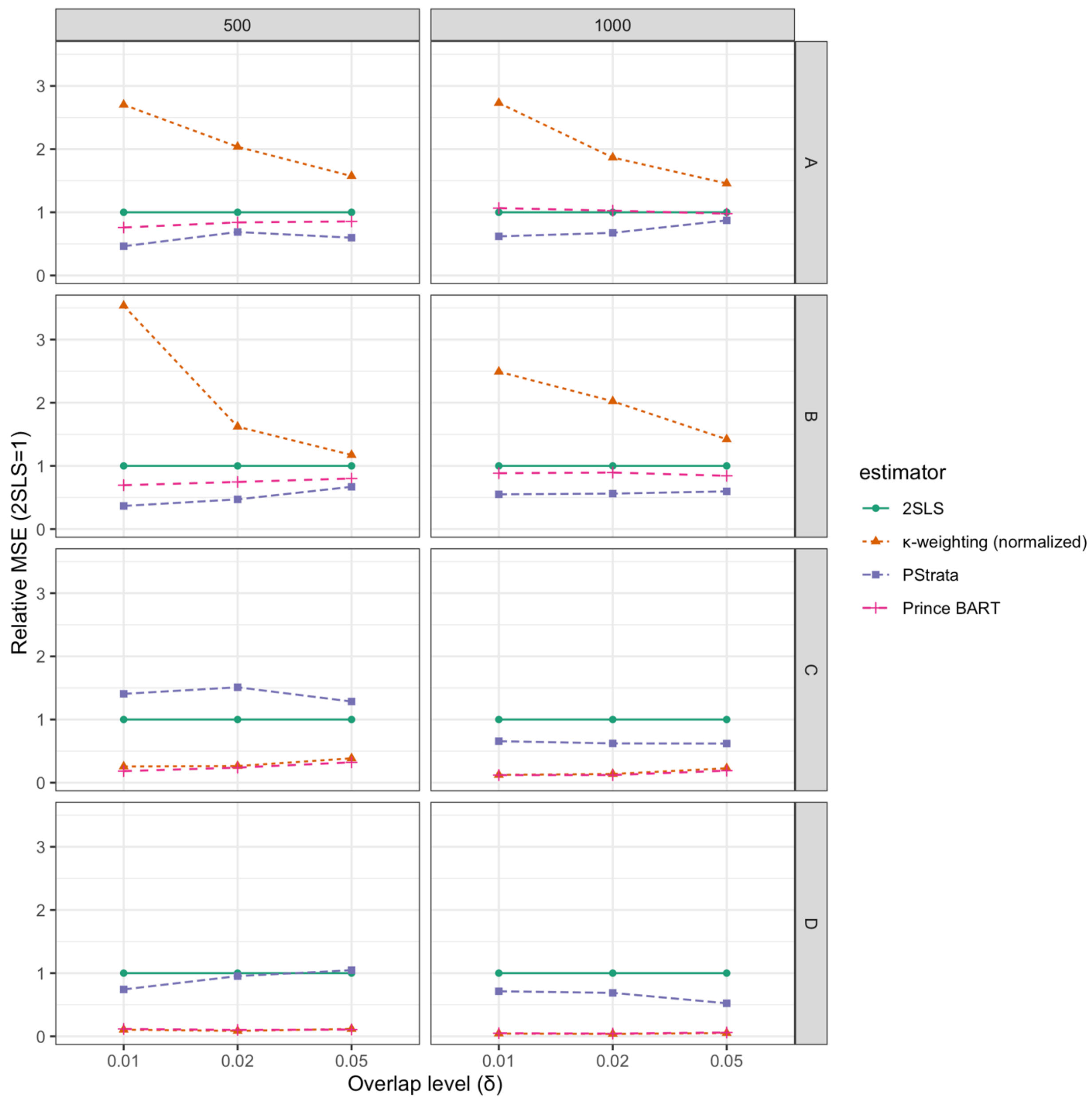

***Figure 2 MSE results for simulation study using 2SLS, κ-weighting (normalized), PStrata, and Prince BART***. *Each panel corresponds to one of the four data-generating designs (rows) and two sample sizes (columns). The x-axis shows the overlap index δ (higher values indicate better propensity score overlap). The y-axis reports mean squared error (MSE) relative to 2SLS (so 2SLS = 1 in every panel). Values below 1 indicate that the estimator has smaller MSE than 2SLS in that scenario.*

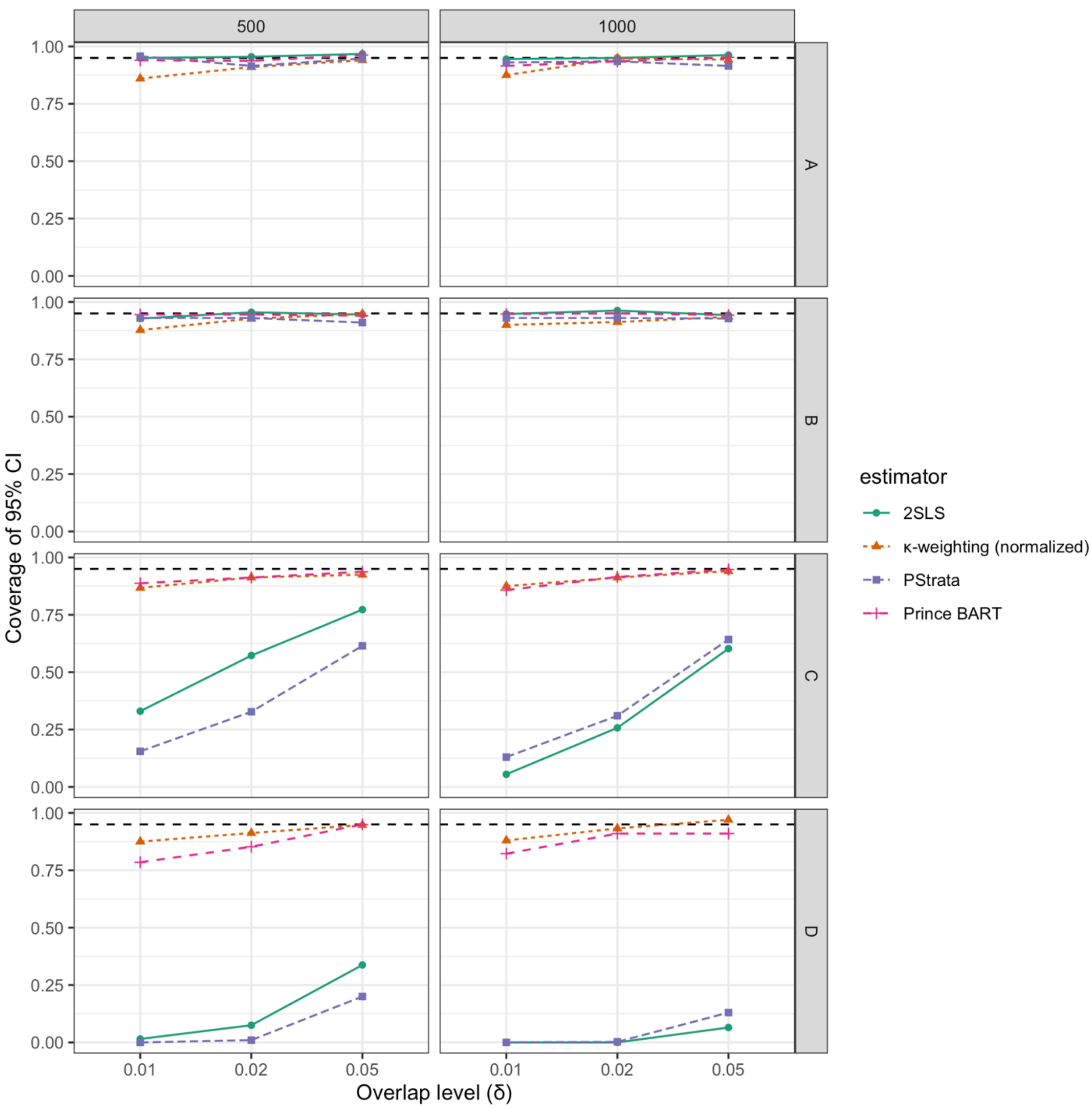

*Figure 3* ***Coverage results for simulation study for 2SLS, κ-weighting (normalized), PStrata, and Prince BART.*** *Each panel corresponds to one of the four designs (rows) and two sample sizes (columns). The x-axis shows the overlap index δ (higher values indicate better propensity score overlap). The y-axis reports empirical coverage of 95% confidence intervals (2SLS and κ-weighting) and 95% posterior credible intervals (PStrata and Prince BART); the dashed line marks the nominal 0.95 level.*

## 6. Application

We use our approach to obtain estimates of the effect of adopting modern contraception on employment in Nigeria and Senegal, using data from MLE and DHS. Throughout this section, "effect of adopting modern contraception" should be interpreted as the causal effect of modern contraceptive use among women whose use is influenced by the FP program (compliers), and its generalizations to broader populations, rather than the direct effect of the program assignment itself. In what follows, we first present results for the source sample (MLE), then for the target populations defined using DHS data. Robustness checks and additional sensitivity analyses are reported separately in the subsequent Section.

### *6.1 Effect of contraceptive use on employment among women affected by an FP program*

We first describe the analysis sample and the estimated complier group. We restricted the MLE source data to women in the intervention and control groups with adequate covariate overlap; this trimming reduces the sample by 17.3% in Nigeria and 0.6% in Senegal (Figure A 2 in Appendix VIII). In the trimmed source samples, an estimated 7.7% of women (sd 0.9%) in Nigeria and 4.8% (sd 0.7%) in Senegal are compliers, i.e., women whose contraceptive use is affected by the FP program rollout timing. Figure A 3 and Tables Table A 10 to Table A 12 in Appendix VIII summarize the baseline characteristics of compliers and target populations.

Table 1summarizes the estimated effects of adopting modern contraception on employment among compliers in the source study ($MATE_C$) for each method. For Nigeria, all four estimators point in the same direction, with positive effects ranging from 0.29 to 0.46. The Prince BART estimate is 0.37 (sd 0.13), lying between the parametric principal-stratification estimate from PStrata (0.29, sd 0.12) and the linear IV and κ–weighting estimates (0.43 and 0.46, sds about 0.22–0.23). Thus, for Nigeria, the methods broadly agree on the sign and order of magnitude of the complier-average effect, with Prince BART and κ–weighting suggesting somewhat larger effects than the parametric PS model and 2SLS, but with overlapping uncertainty intervals.

Agreement across methods is weaker in Senegal. Prince BART and PStrata again yield positive but more modest complier-average effects—0.47 (sd 0.16) and 0.15 (sd 0.14), respectively. The 2SLS and normalized κ–weighting estimates, although positive, fall outside the logical range of –1 to 1 for a probability difference. Because ratio-type IV estimators are not intrinsically bounded, a modest estimated treatment uptake difference can yield estimates outside the admissible range. By contrast, the principal-stratification estimators are defined within probability models and therefore respect the outcome scale.

| Methods | Nigeria | Senegal |
|---|---|---|
| 2SLS | 0.429 (0.223) | 2.389 (0.990)* |
| k-weighting (normalized) | 0.435 (0.222) | 2.321 (0.969)* |
| PStrata | 0.286 (0.120) | 0.217 (0.136) |
| Prince BART | 0.366 (0.133) | 0.473 (0.155) |

*Table 1 Estimated effect of adopting modern contraception on employment among compliers in the source study ($MATE_C$) using alternative methods. Values shown are posterior means (or point estimates) with standard deviations (or standard errors) in parentheses. * Estimate lies outside the logical range of –1 to 1 for a probability difference.*

In both countries, we find substantial evidence of effect heterogeneity among compliers. Using the surrogate shallow tree approach, we partition the Nigeria and Senegal sample of women affected by the FP program into eight subgroups with different estimated effects. Figure 4 depicts the posterior distribution of the difference in the estimated effect between the group with largest and smallest effect in each country. This measure offers strong evidence of effect heterogeneity in Nigeria (the probability that the difference is greater than 0 is 99.8%) and somewhat weaker in Senegal (the probability that the difference is greater than 0 is 88.8%).

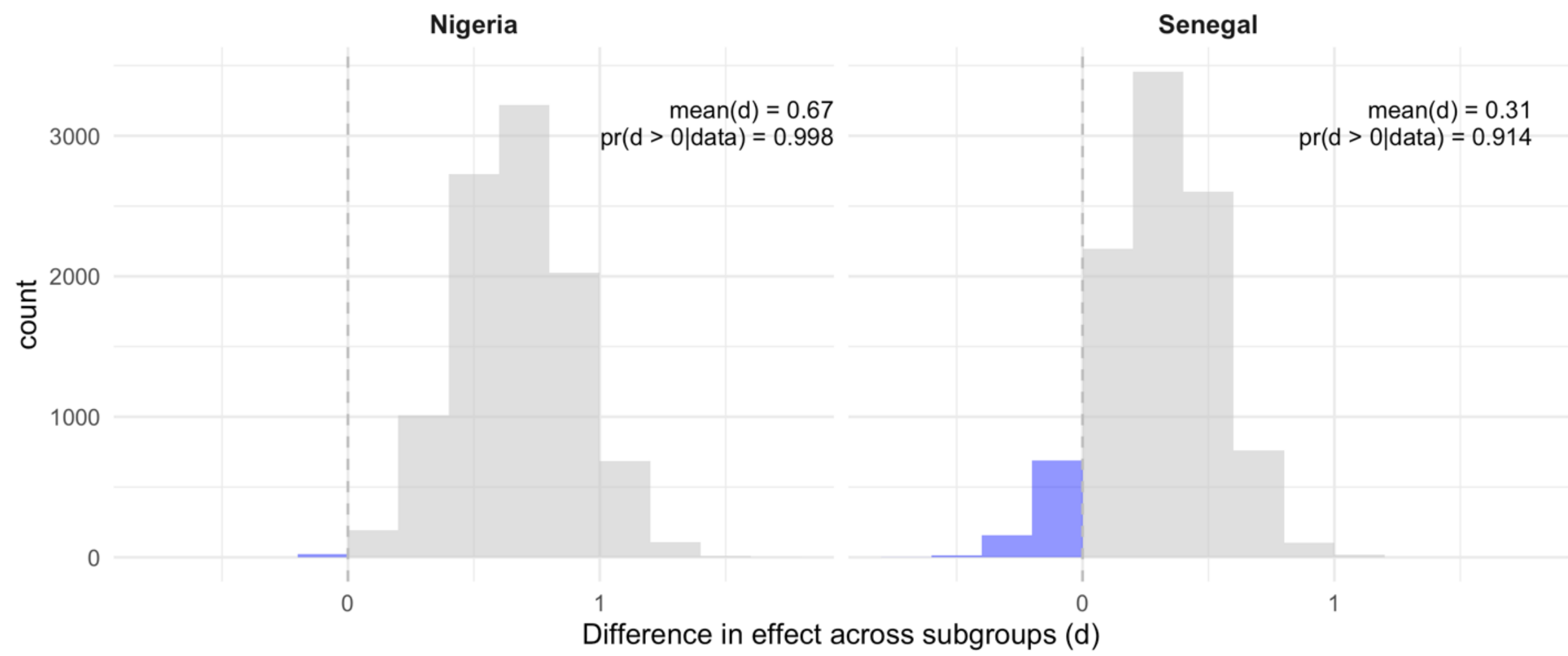

*Figure 4 Posterior distribution of the difference, d, between the effect adopting modern contraception on employment in the segments with smallest and largest effect in Nigeria and Senegal.*

The subgroups identified by the shallow tree approach are defined by a combination of up to three covariates. Figure 5 (panel A) shows the subgroups identified in Nigeria; Senegal results are in Figure 6 (panel A). In Nigeria, we find the lowest effect among the group of women who had not worked during the year prior to baseline, were never married at baseline, and had medium to highest wealth. The largest effect is observed among women who worked during the year prior to baseline, were married at baseline, and had primary education or less. In Senegal, the lowest effect is among women who were not sexually active (which is correlated with marriage), had not worked during the year prior to baseline, and were less than 20 years old. The largest effect is observed among sexually active women who were less than 30 years old and had primary education or less.

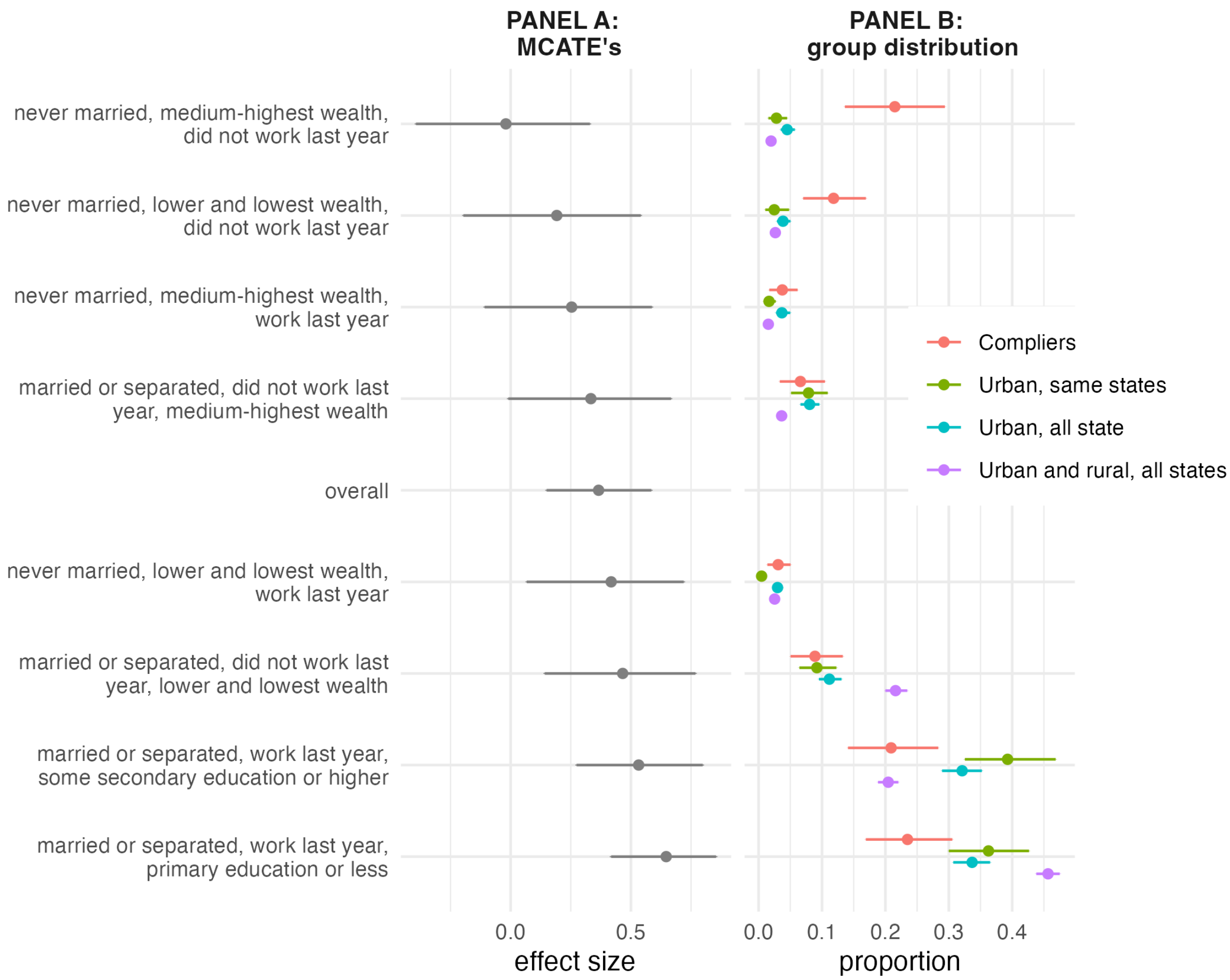

*Figure 5 (A) Effect of contraceptive use on employment by subgroup defined by selected combinations of covariates (MCATEs); (B) proportion of women in the different subgroups for four populations: (1) compliers in the source study, and, based on the DHS information, (2) all women residing in urban areas in the 5 states represented in the source study, (3) all women residing in urban areas, and (4) all women in **Nigeria**. Lines represent 95% credible intervals.*

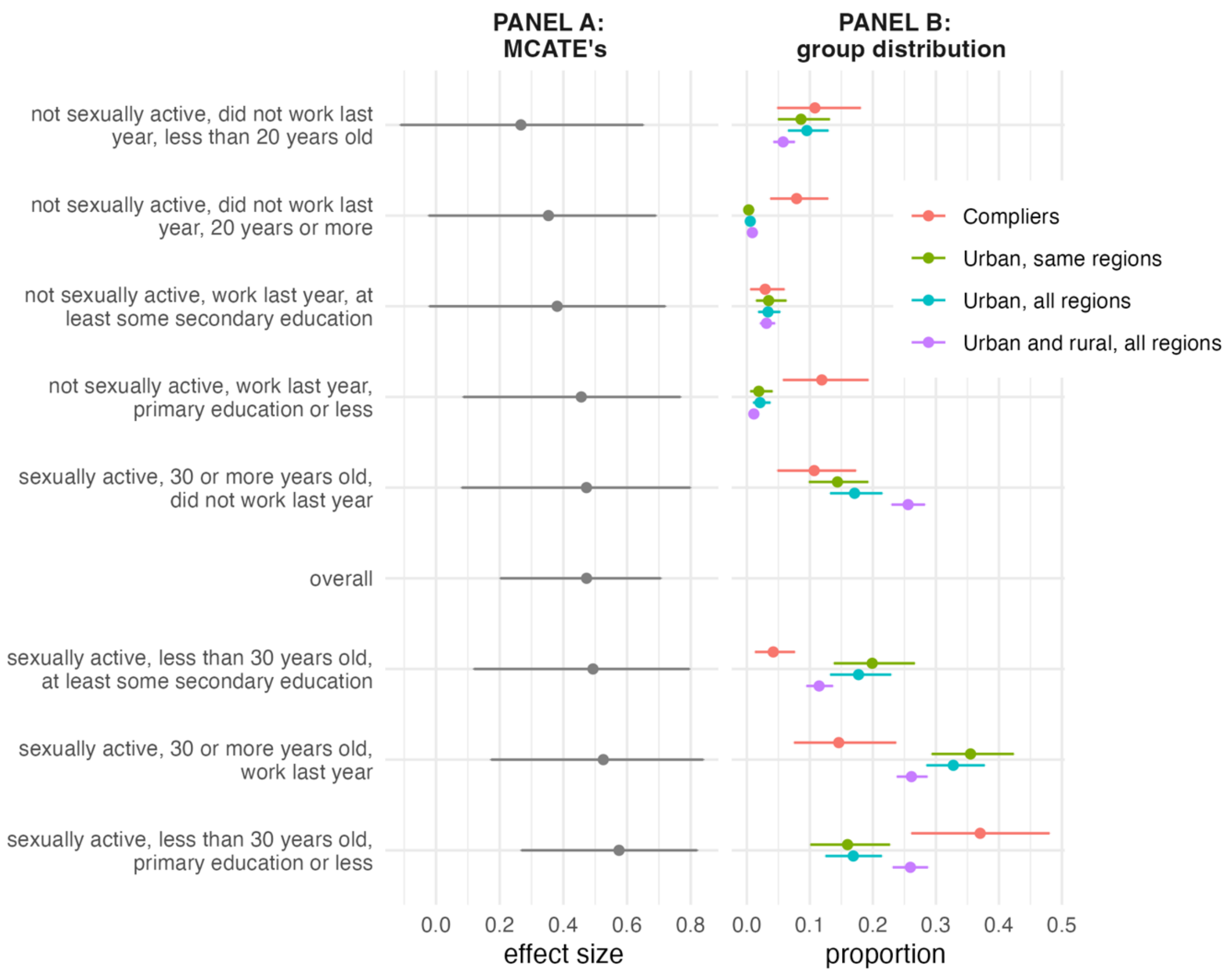

*Figure 6 (A) Effect of contraceptive use on employment by subgroup defined by selected combinations of covariates (MCATEs); (B) proportion of women in the different subgroups for four populations: (1) compliers in the source study, and, based on the DHS information, (2) all women residing in urban areas in the 3 regions in the source study, (3) all women residing in urban areas, and (4) all women in* ***Senegal****. Lines represent 95% credible intervals*

### *6.2 The effect of contraceptive use on employment among Nigerian and Senegalese women*

Figure 5 and Figure 6 (panel B) show the distribution of the subgroups identified in the sample of women affected by the FP program as well as among possible target populations for Nigeria and Senegal, respectively. In the case of Nigeria, the group with the largest effect is underrepresented among compliers while the group with the smallest effect is overrepresented compared to possible target populations. In the case of Senegal, in contrast, the group with the largest effect is overrepresented among compliers. Due to these differences, we do not expect the PATE to be identical to the MATE.

Table 2 reports PATE estimates for different subpopulations of Nigerian and Senegalese women who have not used modern contraception and wish to avoid or delay pregnancy. The DHS questions used to define these populations are listed in Appendix IV. The $MATE_C$ row summarizes the average effect of adopting modern contraception on employment among compliers in the source study, while the PATE rows report the corresponding generalized effects for each target population. For Nigeria, the estimated effects of adopting modern contraception on employment are appreciably larger in these target populations (0.50–0.53) than among compliers in the source sample (0.37), although the 95% confidence intervals largely overlap. Put differently, relative to the source study, these estimates imply that roughly one to two additional women out of every ten would be employed in the target populations if they adopted modern contraception. For Senegal, the PATE estimates are closer to the source complier effect (0.47), consistent with less heterogeneity in the effect and a complier group that is more similar to the target populations (as discussed later in the text). In both Nigeria and Senegal, the estimated uncertainty is larger for PATE than for $MATE_C$. Results obtained under an alternative covariate-imputation procedure are reported in Appendix X and are substantively similar.

| | Nigeria | Senegal |
|---|---|---|
| $MATE_C$ (Prince BART) | 0.366 (0.133) | 0.473 (0.155) |
| PATE | | |
| Urban, same states | 0.518 (0.145) | 0.489 (0.181) |
| Urban, all states | 0.495 (0.143) | 0.483 (0.182) |
| Urban and rural, all states | 0.527 (0.146) | 0.479 (0.195) |

*Table 2 Effect of adopting modern contraception on employment: average effect among compliers in the source study ($MATE_C$, Prince BART) and generalized effects (PATE) for different target populations.*

## 7. Robustness checks

While the assumptions needed for internal and external validity cannot be confirmed, we implement checks that can detect certain departures from key assumptions as well as assess their relevance.

### *7.1 Internal validity - Sensitivity to unobserved city-level confounding*

To assess sensitivity to residual city-level confounding, we posit an unobserved binary predictor $U_i$ that may affect both assignment $Z_i$ and the potential outcomes. We assume that, conditional on $X_i$ and $U_i$, the instrument is as-if random, and that $U_i$ has an additive effect on the outcome on the probit scale. Following Dorie et al. (2016), this leads to a sensitivity analysis indexed by the prevalence difference of $U_i$ across assignment groups and by its predictive strength.

We focus on a worst-case configuration in which $U_i$ is perfectly collinear with assignment, $U_i = Z_i$, as would occur if the confounder operated entirely at the city/program level (i.e., we fix one of the two parameters ats it worth level). In other words, we fix the prevalence imbalance at its maximum value and vary only the outcome effect of $U_i$, yielding a single-parameter sensitivity analysis rather than the two-parameter specification in Dorie et al. Under this simplification, the counterfactual probability of employment at endline, $Y_i^*(1 - Z_i)$, is shifted on the probit scale by a parameter $\kappa$ that captures the strength of unobserved confounding. For interpretability, we write $\kappa = \nu \times \zeta$, where $\nu$ is a benchmark and $\zeta \geq 0$ is a unitless sensitivity parameter. This multiplicative parameterization is similar in spirit to McClean et al. (2024), who also recommend expressing the strength of unobserved confounding as the product of a sensitivity parameter and a problem-specific reference quantity. In our setting, we take $\nu$ to be the residual standard deviation (on the probit scale) of city-specific employment probabilities after adjusting for baseline covariates and contraceptive use, and estimate ν=0.21 in Nigeria and ν=0.16 in Senegal using the calibration procedure described in Appendix III.

Figure 7 shows the estimated complier-specific effect as a function of $\zeta$ (from 0 to 5). In Nigeria, the 90% credible interval includes the null only for $\zeta \gtrsim 3$; in Senegal, only for $\zeta \gtrsim 5$. Thus, unobserved city-level confounding would need to be several times larger than the residual across-city variability unexplained by observed covariates to overturn our qualitative conclusions.

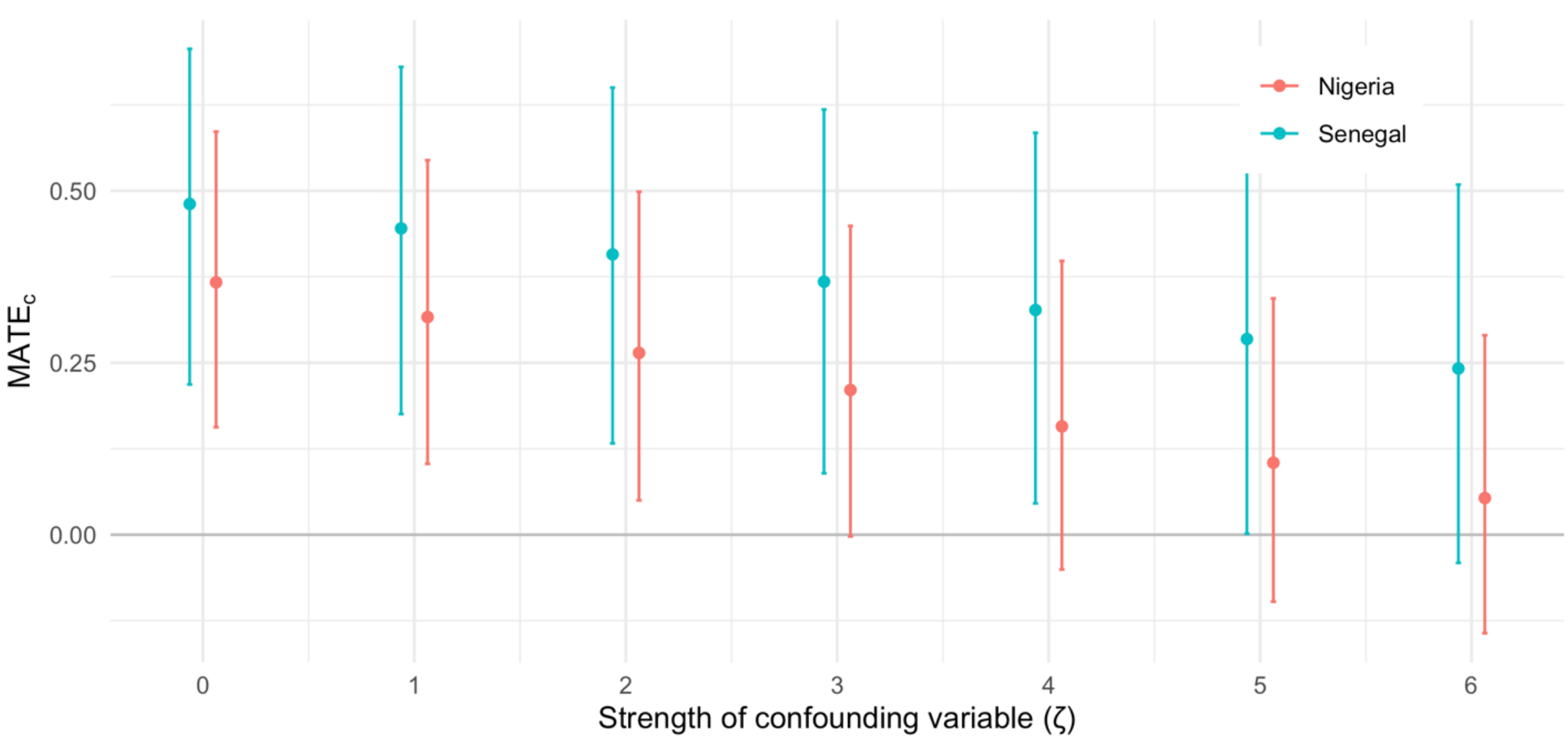

*Figure 7 Sensitivity of the estimated effect to changes in strength of confounder.*

### *7.2 External validity - Sensitivity to violation of support inclusion*

To assess overlap between the source and target populations, we use a scalar *selection score* $s(X_i)$, analogous to a propensity score, defined as

$$s(X_i) = P(\, G_i^* = c, T_i = 0 \mid X_i \,). \tag{13}$$

We decompose this as

$$s(X_i) = P(\, G_i^* = c \mid X_i, T_i = 0\,)\, P(\, T_i = 0 \mid X_i\,), \quad (14)$$

where the first term (the "compliance score") is estimated via BART in the source sample, and the second term is estimated via BART in the stacked MLE–DHS dataset.

We work with a standardized version of the logit of $s(X_i)$, using the mean and standard deviation among compliers in the source sample, and classify target-population women as poorly represented if their standardized selection score falls below the 5th percentile among compliers. Figure 8 displays the distribution of the standardized selection score for (1) compliers in the source sample, (2) urban women in the same states/regions, (3) all urban women, and (4) all women. In Nigeria in particular, a nontrivial share of women in the broader target populations have covariate profiles that are rarely observed among compliers.

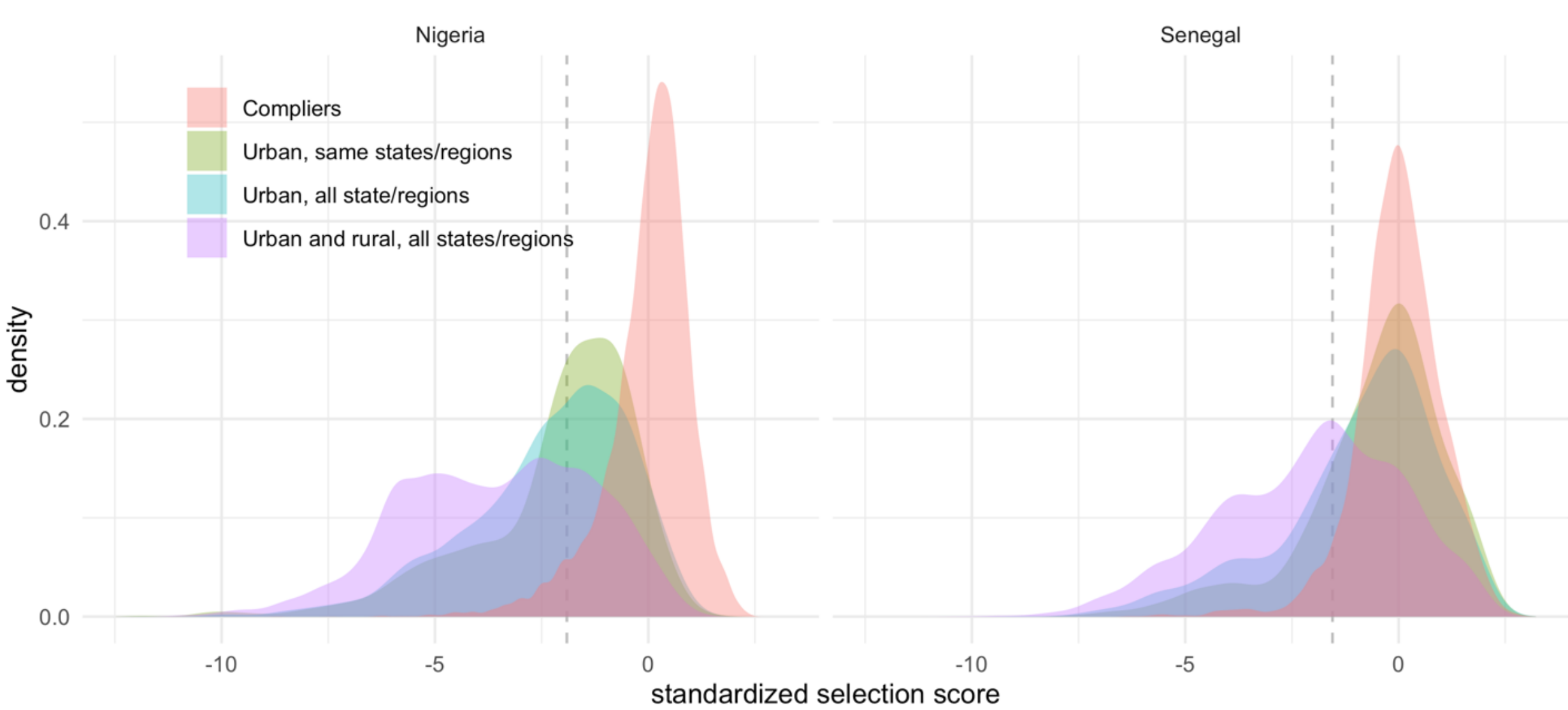

*Figure 8 Estimated standardized selection score as a function of baseline covariates among four populations in Nigeria and Senegal: (1) compliers in the source sample, and (2) all women residing in urban areas in the same states/regions represented in the source sample, (3) all women residing in urban areas, and (4) all women. Dashed grey vertical line represent first percentile in the source sample.*

To gauge the possible impact of this lack of support, we flag target-population women with selection scores below the 5th percentile and recompute the PATE under the extreme assumption that the effect of contraceptive use on employment is zero for these women. The resulting PATEs are reported in Table 3. As we move to broader target populations, the estimated PATE decreases but remains positive and sizable. To drive the PATE to zero, one would need to assume adverse effects of contraceptive use on employment precisely for women whose combinations of baseline characteristics are poorly represented in the source sample.

| | Nigeria | Senegal |
|---|---|---|
| Urban, same states | 0.287 (0.088) | 0.384 (0.139) |
| Urban, all states | 0.218 (0.070) | 0.337 (0.122) |
| Urban and rural, all states | 0.126 (0.037) | 0.211 (0.077) |

*Table 3 PATE for different target populations, assuming null effect for women with combination of baseline characteristics not well represented among compliers in the source study.*

***7.3 External validity - Sensitivity to violation of conditional transportability***

Our main transportability assumption requires that, conditional on $X_i$, the complier-specific effect is the same in the source study and the target population. To examine potential violations due to an unobserved effect modifier $U_i$, we follow the approach of Nie et al. (2021), which yields bounds on the PATE under controlled deviations from conditional transportability.

Let

$$r_i \equiv \frac{f(U_i \mid X_i, T_i = 0)}{f(U_i \mid X_i, G_i^* = c)}, \qquad (15)$$

denote the relative density of $U_i$ in the target and source populations among women with the same observed covariates. Although $U_i$ is unobserved, we can compute sharp bounds on the PATE when $r_i$ is constrained to lie in $[1/\Gamma, \Gamma]$ for a given $\Gamma \geq 1$. We obtain these bounds by solving a linear programming problem. Figure 9 presents the resulting PATE bounds as a function of Γ. For the lower limit of the 95% credible interval for the *minimum* PATE to fall below zero, the maximum density ratio must exceed about 2 in Nigeria and 1.4 in Senegal, indicating that substantial unobserved differences between the source and target populations would be required to reverse our conclusions. For completeness, Appendix XI presents an alternative sensitivity analysis based on a simple parametric model that yields point-identified PATE estimates under parametric assumptions.

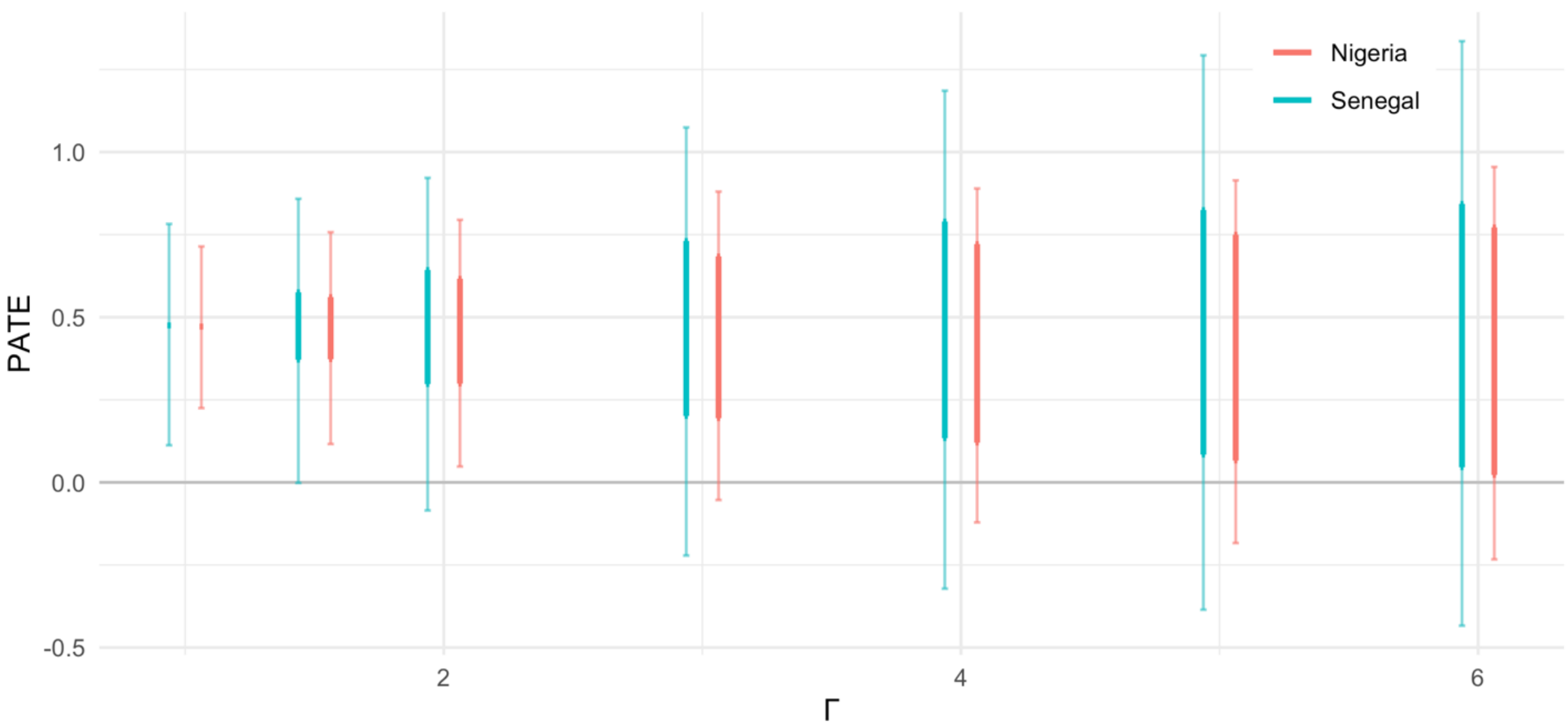

*Figure 9 Sensitivity of the PATE to shift in the covariate distribution (between source and target) induced by an unobserved effect modifier, U, (under a flexible non-parametric model for U). The x-axis quantifies the assumed bound on unobserved distributional shift. The thicker line represents the mean of the posterior distribution of the boundaries, the thinner line represents the 95% credible intervals.*

## 8. Discussion

Estimating the effect of FP on empowerment-related outcomes is difficult because FP uptake and empowerment likely share common causes. In settings where encouragement to use FP can be thought to have been randomly assigned, perhaps after conditioning on covariates, the effect of modern contraceptive use can be identified using principal stratification, among women whose use changes because of the encouragement. These women ("compliers") may differ systematically from the population of interest and those differences may affect the consequences of contraceptive use for empowerment outcomes. In this work, we rely on a Bayesian nonparametric approach to flexibly estimate the conditional average effect in the source sample as well as the covariate distribution in the target population, then combine these to account for shifts in observed covariates. The possible impact of shifts in unobserved covariates is gauged through sensitivity analyses.

We found a strong effect of contraceptive use on employment in two independent applications, Nigeria and Senegal, of roughly the same magnitude. There is also strong evidence of effect heterogeneity, particularly for Nigeria. Unlike Senegal, in Nigeria the average effect of adopting modern contraception in the target population is appreciably higher than in the source study based on point estimates, albeit the 95% credible intervals (CI) largely overlap (more extensive results from Nigeria are presented in Godoy Garraza et al., 2024). In Senegal, the average effect of adopting modern contraception between source and target population are closer. The large differences in Nigeria likely reflect a more heterogeneous population across states and cities in Nigeria as compared to Senegal, where the population is more homogenous across cities and regions. In both study countries, the largest effects were among those women with the least education; this may reflect the important role of contraception on offering these women greater ability to work when they are able to delay or avoid childbearing.

Robustness checks in both countries identified segments of the target population with combinations of covariates underrepresented in the source sample, particularly if the target population included women residing in rural areas. Assuming no effect for these underrepresented segments results in smaller, albeit still sizable, effects in the population. We examined the sensitivity of the results to the presence of an unobserved effect modifier with two different underlying models for the omitted variable. Both approaches suggested that a rather large imbalance in this hypothetical omitted variable would be required to explain away the effect in the population.

Several limitations must be acknowledged. Regarding the use of BART, while BART improves upon commonly used parametric approaches, we note that our approach is subject to assumptions as well, such as normal homoscedastic latent residuals. Limitations in relation to the specific application are discussed in more detail in Godoy Garraza et al., (2024). The main methodological limitation is that we approach the case study as we would a randomized encouragement design trial (Zelen, 1979, 1990), ignoring the fact that assignment is clustered at the city level, with only 6 cities participating. The sensitivity analysis, however, suggests that the results are relatively robust to confounding that could plausibly arise from this source. In addition, confidence in the results benefits from the fact that similar results are obtained from two independent applications. We consider only DHS sampling design (stratified two-stage cluster sampling), which is commonly used in large demographic representative studies. Additional adaptation of the BB design has been proposed in the survey research literature and could be used for causal inference. Regarding the scope, the study does not address how the adoption of contraceptives occurs in the population of interest, if at all. At least for some women, that may require additional FP interventions that could enable or support contraception adoption. However important, we leave that question for future research.

In sum, this study demonstrated an approach to flexibly estimate the effect of FP on empowerment on a subpopulation impacted by a FP program and generalized the results to a broader population of interest, using information from a complex probabilistic sample to estimate the distribution of effect modifiers in the target population. We also showed multiple ways to check the robustness of the results when not all effect modifiers have been addressed. In our view, the main practical value of the proposed approach lies in settings with conditionally independent instruments, where randomization can only be defended within covariate strata and conventional linear IV methods are fragile in the presence of nonlinearities and effect heterogeneity. The findings contribute to stakeholders' understanding of longer-term effects of modern contraceptive use and can help inform decision making regarding family planning policy and investments.

**Acknowledgments**

The authors would like to thank Jocelyn Finlay, Jonathan Bearak, Matt Hamilton, Onikepe Owolabi, Jennifer Seager, and John Stover, as well as Aaron Leor Sarvet, Ted Westling and Nicholas Reich for helpful comments on this work. This paper is a product of the investigator's work within the Family Planning Impact Consortium: a multi-disciplinary partnership between the Guttmacher Institute, African Institute for Development Policy, Avenir Health, Institute for Disease Modeling of the Gates Foundation's Global Health Division, with investigators at the University of Massachusetts Amherst, the University of North Carolina, the George Washington University, and the Institut Supérieur des Sciences de la Population de l'Université Joseph Ki-Zerbo. The Consortium seeks to generate robust estimates of how family planning affects a range of social and economic domains across the life course. Members of the Consortium have developed unique model-based approaches to generating evidence that examines relationships between family planning and empowerment-related variables. Code used for this article is available at https://github.com/AlkemaLab/prince_Bayes

## Funding

This paper was made possible by grants from the Gates Foundation and the Children's Investment Fund Foundation, who support the work of the Family Planning Impact Consortium. The findings and conclusions contained within do not necessarily reflect the positions or policies of the donors. Additional Funder information: Funder: Gates Foundation, Award Number: INV-018349, Grant Recipient: Guttmacher Institute; Funder: Children's Investment Fund Foundation, Award Number: 2012-05769, Grant Recipient: Guttmacher Institute.

## Appendix I: Bayesian mixture model and data augmentation (DA)

Bayesian inference with principal stratification using a data augmentation (DA) was first discussed in Imbens & Rubin (1997). In this appendix we summarize the approach.

Recall that, for each unit in the source sample, we observed a particular realization of 4 random variables $\{Y, Z, W, X\}$. We assume the joint distribution of these variables is governed by a generic parameter $\theta$, with prior distribution $p(\theta)$, conditional on which the random variables are i.i.d. This is quite general, since we have not said anything about the dimensionality of $\theta$. Let $\mathcal{G}(z, w)$ denote the set of principal strata compatible with each combination of $(Z, W)$, e.g., $\mathcal{G}(1,1) = \{c, a\}$. Then the likelihood of the observed data can be written as

$$\prod_{i=1}^{n} P(X_i, Z_i, W_i, Y_i|\theta)$$

$$= \prod_{i=1}^{n} P(X_i|\theta_X)\, P(Z_i|X_i, \theta_Z) \sum_{g \in \mathcal{G}(Z_i, W_i)} P(G_i^* = g|Z_i, X_i, \theta_G)\, P(W_i|G_i^* = g, Z_i, X_i, \theta_W)\, P(Y_i|G_i^* = g, W_i, Z_i, X_i, \theta_Y)$$

$$\propto \prod_{i=1}^{n} \sum_{g \in \mathcal{G}(Z_i, W_i)} P(G_i^* = g|\, X_i, \theta_G) P(Y_i|G_i^* = g, Z_i, X_i, \theta_Y)$$

*(A 1)*

Three terms are absorbed by the proportional sign: (i) the covariate distribution, $P(X_i, \theta_X)$, because the estimands condition on the observed values of the covariates, (ii) the assignment mechanism, $P(Z_i|X_i, \theta_z)$, which is a constant with respect to the outcome; and (iii) the model for the actual "treatment", $P(W_i|G_i^* = g, Z_i, X_i, \theta_W)$, because $W_i$ is a deterministic, on-to-one function of $G_i^*$ and $Z_i$. For the same reason, $W_i$ can be dropped from the conditioning set in $P(Y_i|G_i^* = g, W_i, Z_i, X_i, \theta_W)$. In turn, $Z_i$ can be dropped from $P(G_i^* = g|Z_i, X_i, \theta_G)$ because of unconfoundedness. Unconfoundedness also implies that $P(Y_i|G_i^*, Z_i = z, X_i, \theta_Y) =$ $P\left(Y_i^*\left(Z_i = z, W_i^*(z)\right)|G_i^*, X_i, \theta_Y\right)$, i.e., the outcome model is, equivalently, a model for the potential outcomes.

In summary, we need to specify two models: (i) a principal strata model, denoted by $\pi_g(x) \equiv P(G_i^* = g|X_i, \theta_G)$, and (ii) an outcome model, denoted by $\varpi_{gz}(x) \equiv P(Y_i|G_i^* = g, Z_i, X_i, \theta_Y)$ as well as the prior distribution of the parameters governing these models. We will maintain that the parameters governing these models are distinct and a priori independent of each other and of the parameters governing assignment and covariate distribution.

Given the models and prior for the model parameters, we can approximate the posterior distribution of the causal estimands (i.e., quantities that depend on $\pi$'s and the $\varpi$'s), despite the fact that $G^*$ is missing for the subset units $\{i: Z_i \neq W_i\}$. We use a data augmentation (DA) approach to that end. Let $\tilde{G}$ denote a version of $G^*$ with all unobserved values imputed. A DA algorithm iterates between these two steps,

i. Estimate $\pi$'s and the $\varpi$'s given observed values of $\left(X, Z, W, Y, \tilde{G}\right)$.
ii. Update $\tilde{G}$ (i.e., impute missing values in $G^*$) given observed values $(X, Z, W, Y)$ and current estimates of $\pi$'s and the $\varpi$'s.

The first step is implemented simply as if G was observed; taking its current imputed values as data, we obtain estimates of the latent class and the outcome conditional distributions as a function of the covariates using standard routines. Given estimates $\pi$'s and the $\varpi$'s, we apply Bayes rule to compute the probabilities of class membership, $P(G_i^* = g|X, W, Z, Y, \theta_G)$, conditional on all the observed data including the observed outcome and use it to update $\tilde{G}^*$.

*The DA algorithm*

Let $\tilde{G}$ denote a version of $G^*$ with all unobserved values imputed. A DA algorithm iterates between these two steps,

iii. Estimate the conditional expectations (the $\pi$'s and the $\varpi$'s) with BART given observed values of $(X, Z, W, Y, \tilde{G})$.

iv. Update $\tilde{G}$ (i.e., impute missing values in $G^*$) given observed values $(X, Z, W, Y)$ and the current estimates of $\pi$'s and the $\varpi$'s.

In our implementation, we make extensive use of *dbarts* (Dorie et al., 2024), a BART discrete sampler which facilities incorporating BART within more complex models.

The detailed steps of the DA are as follows:

i. The algorithm needs to be initialized with some values for the missing values in $\tilde{G}$, i.e., for $\{i: Z_i = W_i\}$ .We set the missing values equal to compliers, i.e.,

$$\tilde{G}_i^{(0)} \equiv \begin{cases} a, & if\ Z_i = 0\ \ W_i = 1 \\ n, & if\ Z_i = 1\ \ W_i = 0 \\ c, & elsewhere. \end{cases}$$

For $l = 1, \dots, K$ iterations,

ii. Taken $\tilde{G}_i^{(l-1)}$ as if it were data, we can estimate the latent class probabilities conditional on covariates, $(\pi_c(x), \pi_{a|!c}(x))$ , i.e.,

$$\tilde{\pi}_c^{(l)}(x) \equiv \Pr\left(\tilde{G}_i^{(l-1)} = c | X_i\right) = bart^{c(l)}(x)\ ,$$

$$\tilde{\pi}_{a|!c}^{(l)}(x) \equiv \Pr\left(\tilde{G}_i^{(l-1)} = a | X_i, \tilde{G}_i^{(l-1)} \neq c\right)\ = bart^{a|!c(l)}.$$

Similarly, we can estimate the conditional expectations of the potential outcomes within each latent class, $\left(\varpi_{1c}(x), \varpi_{0c}(x), \varpi_{1a}(x), \varpi_{0n}(x)\right)$, i.e.,

$$\tilde{\varpi}_{zg}^{(l)}(x) \equiv \Pr\left(Y_i = 1 | X_i, \tilde{G}_i^{(l-1)} = g, Z_i = z\right) = bart^{Yzg(l)}(x),$$

for $z = \{0,1\}$ and $g = \{c, n, a\}$. In this step, the BART estimate is given by one posterior sample of the fit.

iii. Taking the current estimated values of the class probabilities and conditional expectations, and given the observed outcome data, we compute the posterior predictive probability $\gamma_{cz} \equiv \mathrm{P}(G_i^* = c | X_i, W_i = Z_i, Y_i)$ for the units where $G_i^*$ is unknown, i.e., $\{i: Z_i = W_i\}$, as follows:

$$\tilde{\gamma}_{c1}{}^{(l)} = \begin{cases} \dfrac{\tilde{\pi}_c^{(l)}(x)\, \tilde{\varpi}_{1c}^{(l)}(x)}{\tilde{\pi}_c^{(l)}(x)\, \tilde{\varpi}_{1c}^{(l)}(x) + \tilde{\pi}_a^{(l)} \tilde{\varpi}_{1a}^{(l)}(x)}, & if\ Y_i = 1 \\[2ex] \dfrac{\tilde{\pi}_c^{(l)}(x)\, \tilde{\varpi}_{1c}^{(l)}(x)}{\tilde{\pi}_c^{(l)}(x)\, \left(1 - \tilde{\varpi}_{1c}^{(l)}(x)\right) + \tilde{\pi}_a^{(l)} \left(1 - \tilde{\varpi}_{1a}^{(l)}(x)\right)}, & if\ Y_i = 0 \end{cases}$$

$$\tilde{\gamma}_{c0}{}^{(l)} = \begin{cases} \dfrac{\tilde{\pi}_c^{(l)}(x)\, \tilde{\varpi}_{0c}^{(l)}(x)}{\tilde{\pi}_c^{(l)}(x)\, \tilde{\varpi}_{0c}^{(l)}(x) + \tilde{\pi}_n^{(l)} \tilde{\varpi}_{1n}^{(l)}(x)}, & if\ Y_i = 1 \\[2ex] \dfrac{\tilde{\pi}_c^{(l)}(x)\, \tilde{\varpi}_{0c}^{(l)}(x)}{\tilde{\pi}_c^{(l)}(x)\, \left(1 - \tilde{\varpi}_{0c}^{(l)}(x)\right) + \tilde{\pi}_n^{(l)} \left(1 - \tilde{\varpi}_{0n}^{(l)}(x)\right)}, & if\ Y_i = 0 \end{cases}$$

iv. Based on these posterior probabilities, impute new values for $G_i^*$,

$$\left(\tilde{G}_i^{(l)} |\, Z_i = W_i = z\right) \sim Bernoulli\left(\tilde{\gamma}_{cz}^{(l)}\right)$$

for $z = 0,1$.

We run 20 chains of 4000 iterations, discarding the first 2000, and thinning afterwards to store 10,000 samples. In our application this ensures $\hat{R} \leq 1.03$ and effective sample size of several hundreds (Vehtari et al., 2020).

**Appendix II: Bayesian Additive Regression Trees (BART)**

We use BART to flexible model latent class membership probabilities as a function of observed covariates as well as expected outcome conditional on class membership and covariates. BART has been previously used for causal inference (Hill, 2011; Dorie et al., 2019; Hahn et al., 2020). Hill et al., (2020) provides a recent review of the method. In this appendix we discuss the approach used for our application in more detail.

### *1. The BART approach*

In the absence of a parametric model, a natural strategy to estimate an unknown regression function is by partitioning the covariate space into cells and then estimating the function locally from available observations within each cell. This is the basic idea of tree-based approaches.

While intuitive and easy to interpret, models based on a single tree (i.e., a single set of splitting rules resulting on a single set of partitions) are known to offer only poor predictive performance. For starters, there is the lack of smoothness. At least in its basic flavor, the same prediction (the average outcome in that region) applies to the entire covariate region, i.e., the tree is a step function.

Ensembles of tree, on the other hand, can perform substantially better even if they are no longer that easy to interpret or represented graphically. A random forest, for example, averages the prediction of many trees fitted to random subsamples of units using only random subset of predictors. Gradient boosting adds up predictions from multiple trees, fitted recursively to the residuals of the previous fit, each one induced to "underfit" the data by a penalization parameter.

BART (Chipman et al., 2007, 2010, onwards CGM ) is an ensemble of trees, typically between 50 and 200 of them. As in gradient boosting, each tree is constrained to be a "weak learner", explaining only a part not already explained by the others. Rather than using a penalization parameter, BART avoids overfitting by using prior distributions that favor small trees with predictions for its terminal nodes not far from the global average. Because a probabilistic model is used for this forest, BART results in a posterior distribution for the estimated regression function of interest.

Two essential components of BART are the sum-of-trees model and the regularization prior. We will first describe these two components focusing on a continuous outcome and then describe the modification for binary outcomes, as in our application.

### *2. The sum-of-trees model*

Let $T$ denote a binary tree consisting of a set of rules segmenting the predictor space into non-overlapping regions, say $R_1, \ldots, R_b$. Binary trees admit only certain types of rules, i.e., binary splits of the predictor space of the form $\{x \in A\}$ vs $\{x \notin A\}$ where $A$ is a subset of the range of x. Each split is referred as an internal node, while the resulting partitions are referred as terminal nodes or "leaves". The set of splitting rules used to segment the predictor space can be summarized in a tree diagram (typically drawn upside down, in the sense that the leaves are at the bottom of the tree).

Let $M = (\mu_1, \ldots., \mu_b)$ denote the set of parameters for tree *T*. Given $(T, M)$, a regression tree is a step function, $h(x; T, M)$, that assign the value $\mu_k$ whenever $x \in R_k$. BART approximates the unknown function $f(x) = E(Y|x)$ , i.e., the conditional expectation of the response given a set of predictors, as a sum of m of these step functions, i.e.,

$$f(x) = \sum_j^m h(x; T_j, M_j) = \sum_j^m \sum_k^{b^j} 1\left(x \in R_k^j\right) \mu_k^j.$$

If a single tree were to be used to approximate $f(x)$, the parameters of the terminal nodes of the tree, the $\mu's$, would correspond to the conditional expectation for each region. When, instead, an ensemble of trees is used, each one contributes only a part of this expectation, the part that remains unexplained by the rest of the trees in the ensemble.

### 3. *A regularization Prior*

A complete model specification requires postulating a prior over each of the parameters of the sum-of-trees model, namely, $\{(T_1, M_1), \dots, (T_m, M_m)\}$.[3] This is a large number of parameters, [4] but the task can be simplified by assuming that, a priori, the distribution of all trees, and of the terminal node parameters within each tree, are independent and the same.[5] In such scenario, there

---

[3] We may also have to specify priors for additional parameters that arise in the data generating mechanism, such as $\sigma$ if the outcome is continuous. We omit that discussion here given that it does not apply to our case.

[4] For example, for m=200 and assuming 3 terminal nodes per tree on average (i.e., 2 splitting rules and 3 terminal nodes parameters) the entire model would contain 1,000 parameters. The actual number of parameters is not prespecified, not even for fixed m, since the tree complexity depends on the data (the prior is posed on the tree-generating process).

[5] In such scenario, the prior for the sum of trees can be factorized as

$$p(T_1, M_1, \dots, T_m, M_m) = \prod_j p(T_j, M_j) = \prod_j p(M_j|T_j)p(T_j),$$

and further,

$$p(M_j|T_j) = \prod_i p(\mu_{ij}|T_j)$$

where $\mu_{ij} \in M_j$ .The independence restriction simplifies the prior specification problem to the specification of the form for just $p(T_j)$, and $p(\mu_{ij}|T_j)$. If a priori the distributions are the same, we can drop the indices.

is only need to specify the distribution of a single tree, $p(T)$, and a single terminal node parameter, $p(T)$.

Priors for the splitting rule $p(T)$

Instead of specifying a closed-form expression for the tree prior, $p(T)$, the distribution is specified implicitly by a tree-generating stochastic process, a branching process. Each realization of such a process can be considered as a random draw from this implicit prior distribution.

The tree-generating process is specified by two aspects: (i) the probability that a node at depth $d$ (for $d = 0, 1, ...$) is nonterminal (equivalently, the probability that the node is split); and (ii) the distribution on the splitting rule if the node is split.

CGM proposed specifying the probability that a node at depth $d$ is nonterminal as $\alpha(1 + d)^{-\beta}$, with $\alpha \in (0,1)$ and $\beta \in [0, \infty)$. Under this specification the probability of a node being split decrease with depth, and more so for large $\beta$. For example, with the choice $(\alpha, \beta) = (.95, 2)$, which is CGM's proposed default, trees with 1, 2, 3, 4 and ≥5 terminal nodes receive prior probability of 0.05, 0.55, 0.28, 0.09 and 0.03, respectively.

If the node is split, the splitting rule encompasses a choice of both a predictor and a cut-point to split. CGM propose choosing the predictor uniformly from the available predictors, and the cut-point uniformly from the available observed values of the selected predictor (or choosing the subset of categories uniformly from the set of available subsets if the predictor is categorical). Alternative priors have been suggested to induce sparsity such as "spike-and-tree" (Ročková & Van Der Pas, 2020) or conditionally-conjugate Dirichlet priors (Linero, 2018). In our application, we stick to the uniform prior set up.

Priors on the terminal value $p(\mu|T)$

For each terminal node within each tree a conjugate normal distribution is used, i.e.,

$$p(\mu|T) \sim N(\mu_\mu, \sigma_\mu^2).$$

CGM proposed to set the values of the hyperparameters $(\mu_\mu, \sigma_\mu)$, using information from the sample. Under the sum-of-trees model, the induced prior for $E(Y|x)$ is $N(m\mu_\mu, m\sigma_\mu^2)$. [6] It is reasonable to expect that $E(Y|x)$ is between the observed minimum and maximum of Y in the data. We can choose $(m\mu_\mu, m\sigma_\mu^2)$ so that $N(m\mu_\mu, m\sigma_\mu^2)$ assigns a substantial probability to that interval. For instance, with over 95% probability, $N(m\mu_\mu, m\sigma_\mu^2)$ will be in the range $(m\mu_\mu \pm k\sqrt{m}\sigma_\mu)$ for $k = 2$. Thus, with observed continuous outcomes, we can set $\mu_\mu = \frac{\bar{y}}{m}$, and $\sigma_\mu = \frac{y_{max} - y_{min}}{k2\sqrt{m}}$, where $\bar{y}$, $y_{max}$ and $y_{min}$ are the sample mean, minimum and maximum values, respectively. [7]

This prior has the effect of shrinking the $\mu$'s towards $\frac{1}{m}$ of the overall average (and shrinking $f(x) = E(Y|x)$ towards $\bar{y}$ ). As $k$ and/or the number of trees $m$ is increased, this prior will become tighter and apply greater shrinkage. This prevents overfitting as the number of trees increases. This choice of a conjugate prior has subsequent computational advantages.[8]

---

[6] Linero & Yang (2018) asserts this prior converge to a Gaussian process as m→∞.

[7] For convenience, CGM suggested shifting and rescaling Y, so that the minimum, mean, and maximum are (-.5, 0, .5), respectively.

### *4. BART with binary outcomes*

An extension to binary outcomes was suggested in CGM's original articles based on the probit model, i.e.,

$$p(x) \equiv \Pr(Y = 1|x) = \Phi\left(\sum_j h_j(x)\right)$$

where $\Phi(.)$ is the standard normal cdf. There is an equivalent formulation in terms of a latent variable, $Z^*$, which is only observed to cross zero, i.e.,

$$Y = 1\{Z^* > 0\},$$

$$Z^* = \sum_j h_j(x) + \epsilon$$

where $\epsilon$ follows a standard normal distribution. This formulation makes the connection with the continuous case more evident. It is reasonable to expect that $p(x)$ to be within the interval $\big(\Phi(-3), \Phi(3)\big)$. [9] The prior for the terminal node parameters can be chosen so there is a priori high probability for that event. Setting $\sigma_\mu = \frac{3-(-3)}{2k\sqrt{m}} = \frac{3}{k\sqrt{m}}$ and choosing $k = 2$, CGM suggested default, there is a priori 95% probability that $p(x)$ within intended range. We can shrink towards a value other than .5 by introducing an offset, say $\Phi^{-1}(p_0)$.

---

[8] In particular, the likelihood of a tree $L(T) \equiv p(y|x,T) = p(y|x,\mu,T)p(T)d\mu$ , can be obtained analytically. Similarly, we can quickly obtain the posterior distribution of a tree up to a normalizing constant, i.e., $p(T|y,x) \propto L(T)p(T)$. This offers a means to quickly compare the posterior probability of two trees.

[9] Unlike the case with the observed continuous outcome, the maximum and minimum of $Z^*$ are not observed and could in principle be infinity, which is not useful to set the priors.

### *5. Bayesian backfitting MCMC algorithm*

The Bayesian backfitting algorithm reduces estimation of the entire posterior

$$p\big((T_1, M_1), \ldots, (T_m, M_m)|y\big)$$

to the much simpler problem of estimating a single tree many times.

Backfitting is a common strategy in the context of frequentist estimation of generalized additive models. Such models express the response variable as a sum of (typically nonlinear) functions of the predictor variables. Estimation of the entire model can proceed by repeatedly updating the fit for each function separately, holding the others fixed, and focusing on the partial residuals. Hastie & Tibshirani (2000) proposed that, by adding appropriate noise at each iteration, a new realization of the current function can be obtained, equivalent to Gibbs sampling from the appropriately defined Bayesian model. The algorithm to fit BART uses a version of this procedure.

For a fixed number of trees m, BART uses an iterative backfitting algorithm to cycle over and over through the m trees. At each iteration, rather than fitting a fresh tree to the partial residuals, BART randomly chooses a perturbation to the tree from the previous iteration from a set of possible perturbations, favoring ones that improve the fit to the partial residuals.[10] Chipman et al., (1998) proposed to consider four possible perturbations: splitting a current leaf into two new leaves (grow), collapsing adjacent leaves back into a single leaf (prune), reassigning the decision rule attached to an interior node (change), or swapping the decision rules assigned to two interior nodes (swap). After the tree is modified, the other parameters (the $\mu's$ in our application) are updated by sampling from their conditional distribution.

[10] Based on the ratio of the posterior probabilities of the trees.

In the case of binary outcomes, the backfitting algorithm is not fitted to the observed binary outcome but to the underlying latent variable, $Z^*$, which therefore needs to be imputed at each iteration.

## *6. Summarizing BART results*

While BART is more flexible than logistic regression, it is also less easy to interpret. A general strategy to summarize complex "black box" models is to fit simpler, surrogate models (Molnar et al., 2020). We use variations of this strategy, termed surrogate deep and shallow tree, respectively, to: (i) identify relevant predictors of latent class membership and of outcome, conditional on class membership, (ii) identify combinations of predictors defining segments with relatively homogenous CATEs.

### Surrogate "deep" trees to identify relevant predictors

Carvalho et al. (2020) suggest the use of deep trees to identify relevant predictors. In this approach, the goal is to approximate BART predictions with a flexible function of only a handful of the covariates. At this point, there is no interest in learning or understanding the approximating function itself. Instead, we care about keeping it as flexible as possible, to reflect the fact that predictors may be relevant in different ways (e.g., by interacting with other predictors).

Let $\hat{y}$ denote the predicted values (the posterior mean or median) from BART for the response variable $y$ based on the entire set of covariates $X$ of dimension $p$. Consider a subset of $X$, of dimension $s < p$, say $Q$. We fit a single regression tree to $\hat{y}$ as a function of $Q$ using a standard algorithm (Breiman et al., 1984; Therneau & Atkinson, 2022), but letting the tree grow without constraints. We obtain new fitted value say $\ddot{y}$ based on this deep surrogate regression tree. These are predictions of the fitted values from BART (not predictions of the outcome itself, $y$) based on only a subset of the predictors. We assess how close is $\ddot{y}$ to $\hat{y}$ using Person $R^2$.

Initially, we consider all possible subsets of size one and chose the subset with the larger $R^2$. Starting from that subset (i.e., the single best predictor), we use a stepwise forward algorithm to consider subsets of covariates of increasing size. The $R^2$ tends to increase fast initially and slows down as the number of predictors included grows larger. We stop when an additional predictor will not increase $R^2$ by more than 1%.

The procedure is useful to identify a handful of relevant predictors without restricting the functional form of the relationship between these predictors and BART fitted values. No claim is made that the subset identified is the only possible subset of relevant predictors.

Surrogate "shallow" tree to identify relevant segments

To learn how the set of predictors identified maybe important, we rely on a second regression tree. The response variable is again $\hat{y}$ , the fitted value from BART, which we regress on the set of relevant predictors identified in the previous step, say $Q^*$. Unlike the first step, we now constrain the tree to a maximum depth of 3. This constraint reflects that the priority in this step is to understand the relationship, more than predict the fitted values with maximum accuracy. A similar procedure is used in  Logan et al., (2019), to examine effect heterogeneity - other examples include J. Hill & Su (2013); Hahn et al. (2020);  and Chen et al. (2024).

**Appendix III: Calibration of $\nu$ for the sensitivity analysis**

In the robustness checks, the strength of unobserved city-level confounding enters the sensitivity analysis through the parameter $\kappa = \nu \times \zeta$. To aid interpretation, we express $\kappa$ as a multiple $\zeta$ of a benchmark $\nu$ that captures the expected magnitude of residual across-city variability in employment. We obtain $\nu$ using a simple two-step calibration procedure.

First, we flexibly estimate the probability of employment at endline as a function of baseline covariates and contraceptive use using probit BART,

$$P(Y_i = 1 \mid X_i, W_i) = \text{bart}(X_i, W_i).$$

Second, we fit a probit regression with the BART prediction as an offset and city indicators as the only regressors,

$$P\big( Y_i = 1 \mid \bar{b}_i, \text{city}_i \big) = \Phi\left( \bar{b}_i + \sum_c \alpha_c \, \mathbf{1}\big(\text{city}_i = c\big) \right),$$

where $\bar{b}_i \equiv \Phi^{-1}\big(\text{bart}(X_i, W_i)\big)$ and $c$ indexes cities. Variation in the estimated city intercepts $\{\alpha_c\}$ reflects residual differences across cities not explained by observed covariates or contraceptive use. We set $\nu$ equal to the standard deviation of these intercepts, obtaining $\nu = 0.21$ in Nigeria and $\nu = 0.16$ in Senegal.

## Appendix IV: Target population(s)

| Variable | Definition | Population | | |
|---|---|---|---|---|
| | | Urban, same states | Urban, all states | Rural and urban, all states |
| Ever use modern contraception | V302A “ever used anything or tried to delay or avoid getting pregnant” > 0 “no” AND<br>V301 “knowledge of any method” = 3 “knows modern method” | F | | |
| Currently using modern contraception | V313 “current use by method type” = 3 “modern method” | F | | |
| Wish to delay or space pregnancy | V605 “desire for more children” in (“wants after 2+ years”, “wants no more sterilized (respondent or partner)”) | T | | |
| Urban | V025 “Type of place of residence” = “urban” | T | | * |
| In 5 states (Nigeria)/ In 3 regions (Senegal) | SSTATE in (Abuja; Edo; Kwara; Kaduna; Oyo)/<br>v024 in (Dakar, Kaolack, Thiès) | T | * | * |

*Figure A 1 Definition of target population based on DHS Nigeria dataset NGIR7BFL. “T”: TRUE, “F”: FALSE, “*”: either TRUE or FALSE.*

## Appendix V: Bayesian Bootstrap to estimate the distribution of the covariates in the target population

To estimate the covariate distribution int the target population, $P_{X|T=1}(x)$, we use Bayesian bootstrap (Rubin, 1981) adapted to complex sampling. We provide additional detail of the approach in this appendix.

We will assume that X can only take a finite number of distinct values, albeit potentially a very large number. In this setting, the goal becomes to estimate the probabilities associated with each of these values based on the target population sample.

Introducing some notation, let $d = (d_1, \dots, d_o)$ be the distinct values of X. Because X is multidimensional, each $d_j$ is a vector of the same dimension - in our application, for example, 47. Let $\varphi = (\varphi_1, \dots, \varphi_o)$ be the associated vector of probabilities, such that

$$p_{X|T=1}(X = d_j|\varphi) = \varphi_j, \quad and \sum_j \varphi_j = 1.$$

*(A 2)*

If the population survey was based on SRS, the observed data would consist of combinations $(d_j, n_j)$ where $n_j$ refers to the number of times the distinct value $d_j$ was observed in the data. In such case, the observed counts would follow a multinomial distribution, and the likelihood function is proportional to,

$$L(\varphi_1, \dots, \varphi_o|n_1, \dots, n_o) \propto \prod_{j=1}^{o} \varphi_j^{n_j}.$$

*(A 3)*

To obtain the posterior distribution of the $\varphi's$, and thus of $P_{X|T=1}(x)$, we need to specify a prior for the $\varphi's$. The Bayesian bootstrap is obtained by posing an improper Dirichlet proportional to $\prod_{j=1}^{k} \varphi_j^{-1}$ (sometimes termed Haldane prior). This approach was termed Bayesian Bootstrap by Rubin (1981) who also describes a straightforward algorithm to sample from the $\varphi's$ posterior. While postulating a finite support for X is not particularly restrictive, the prior does restrict analysis to the observed support, in the same way as the frequentist bootstrap.

Due to the complex sampling design, the observations from DHS cannot be considered an iid sample from X. We consider a BB implementation to account for the complex DHS sampling design, as detail in the body of the article. While similar BB procedures have been proposed (Makela et al., 2018; Rao & Wu, 2010; Zangeneh & Little, 2015) we are unaware of a customary name for the procedure. We termed it the "scaled" or modified BB.

## Appendix VI: Additional Simulation Study Results

Here we report detailed Monte Carlo results for all designs considered in the main text. For each of the four data-generating designs and for every combination of sample size $n$ and overlap index $\delta$ (the minimum and 1–maximum propensity score, where larger $\delta$ indicates better overlap), Tables A.1–A.4 present the performance of the three estimators—2SLS, $\kappa$-weighting (normalized), and Prince BART. Rows correspond to $(n, \delta)$–measure combinations, and columns to estimators. For each scenario we report the mean squared error relative to 2SLS (so that 2SLS = 1), the absolute bias, and the empirical coverage of nominal 95% intervals (confidence intervals for 2SLS and $\kappa$-weighting, posterior credible intervals for Prince BART). All quantities are based on 400 Monte Carlo replications and complement the graphical summaries of relative MSE and coverage shown in Figures 1 and 2 of the main text.

| n | delta | measure | 2SLS | $\kappa$-weighting (normalized) | PStrata | Prince BART |
|---|---|---|---|---|---|---|
| 500 | 0.01 | MSE | 1.000 | 2.704 | 0.462 | 0.759 |
| 500 | 0.01 | \|Bias\| | 0.010 | 0.025 | 0.014 | 0.037 |
| 500 | 0.01 | Coverage | 0.950 | 0.860 | 0.958 | 0.940 |
| 500 | 0.02 | MSE | 1.000 | 2.037 | 0.688 | 0.840 |
| 500 | 0.02 | \|Bias\| | 0.000 | 0.007 | 0.022 | 0.033 |
| 500 | 0.02 | Coverage | 0.955 | 0.910 | 0.915 | 0.938 |
| 500 | 0.05 | MSE | 1.000 | 1.573 | 0.598 | 0.855 |
| 500 | 0.05 | \|Bias\| | 0.001 | 0.003 | 0.002 | 0.022 |
| 500 | 0.05 | Coverage | 0.968 | 0.940 | 0.948 | 0.963 |
| 1000 | 0.01 | MSE | 1.000 | 2.729 | 0.619 | 1.066 |
| 1000 | 0.01 | \|Bias\| | 0.007 | 0.013 | 0.002 | 0.030 |
| 1000 | 0.01 | Coverage | 0.943 | 0.870 | 0.930 | 0.915 |
| 1000 | 0.02 | MSE | 1.000 | 1.868 | 0.674 | 1.025 |
| 1000 | 0.02 | \|Bias\| | 0.001 | 0.003 | 0.005 | 0.021 |
| 1000 | 0.02 | Coverage | 0.948 | 0.948 | 0.935 | 0.935 |
| 1000 | 0.05 | MSE | 1.000 | 1.462 | 0.872 | 0.978 |
| 1000 | 0.05 | \|Bias\| | 0.007 | 0.002 | 0.002 | 0.000 |
| 1000 | 0.05 | Coverage | 0.965 | 0.940 | 0.915 | 0.953 |

*Table A 1 Monte Carlo performance of 2SLS, κ-weighting (normalized), PStrata, and Prince BART in Design A. Rows correspond to combinations of sample size $n$, overlap index $\delta$, and performance measure (relative MSE, absolute bias, and coverage). Columns correspond to the three estimators. Relative MSE is normalized so that 2SLS has value 1 in each $(n, \delta)$ scenario. Coverage refers to nominal 95% confidence intervals for 2SLS and κ-weighting and 95% posterior credible intervals for Prince BART. Results are based on 400 Monte Carlo replications.*

| n | delta | measure | 2SLS | $\kappa$-weighting (normalized) | PStrata | Prince BART |
|---|---|---|---|---|---|---|
| 500 | 0.01 | MSE | 1.000 | 3.540 | 0.366 | 0.695 |
| 500 | 0.01 | \|Bias\| | 0.024 | 0.031 | 0.011 | 0.022 |
| 500 | 0.01 | Coverage | 0.930 | 0.877 | 0.930 | 0.945 |
| 500 | 0.02 | MSE | 1.000 | 1.618 | 0.470 | 0.746 |
| 500 | 0.02 | \|Bias\| | 0.027 | 0.002 | 0.014 | 0.017 |
| 500 | 0.02 | Coverage | 0.955 | 0.927 | 0.930 | 0.945 |
| 500 | 0.05 | MSE | 1.000 | 1.172 | 0.669 | 0.801 |
| 500 | 0.05 | \|Bias\| | 0.019 | 0.006 | 0.011 | 0.016 |
| 500 | 0.05 | Coverage | 0.945 | 0.948 | 0.910 | 0.940 |
| 1000 | 0.01 | MSE | 1.000 | 2.500 | 0.550 | 0.884 |
| 1000 | 0.01 | \|Bias\| | 0.027 | 0.004 | 0.012 | 0.016 |
| 1000 | 0.01 | Coverage | 0.948 | 0.900 | 0.930 | 0.948 |
| 1000 | 0.02 | MSE | 1.000 | 2.032 | 0.561 | 0.895 |
| 1000 | 0.02 | \|Bias\| | 0.026 | 0.012 | 0.013 | 0.004 |
| 1000 | 0.02 | Coverage | 0.963 | 0.912 | 0.930 | 0.950 |
| 1000 | 0.05 | MSE | 1.000 | 1.466 | 0.597 | 0.843 |
| 1000 | 0.05 | \|Bias\| | 0.014 | 0.007 | 0.000 | 0.005 |
| 1000 | 0.05 | Coverage | 0.950 | 0.930 | 0.927 | 0.943 |

*Table A 2 Monte Carlo performance of 2SLS, κ-weighting (normalized), PStrata, and Prince BART in Design B. Rows correspond to combinations of sample size $n$, overlap index $\delta$, and performance measure (relative MSE, absolute bias, and coverage). Columns correspond to the three estimators. Relative MSE is normalized so that 2SLS has value 1 in each $(n, \delta)$ scenario. Coverage refers to nominal 95% confidence intervals for 2SLS and κ-weighting and 95% posterior credible intervals for Prince BART. Results are based on 400 Monte Carlo replications.*

| n | delta | measure | 2SLS | $\kappa$-weighting (normalized) | PStrata | Prince BART |
|---|---|---|---|---|---|---|
| 500 | 0.01 | MSE | 1.000 | 0.261 | 1.407 | 0.183 |
| 500 | 0.01 | \|Bias\| | 0.293 | 0.037 | 0.358 | 0.101 |
| 500 | 0.01 | Coverage | 0.332 | 0.860 | 0.155 | 0.887 |
| 500 | 0.02 | MSE | 1.000 | 0.259 | 1.511 | 0.238 |
| 500 | 0.02 | \|Bias\| | 0.208 | 0.022 | 0.262 | 0.068 |
| 500 | 0.02 | Coverage | 0.565 | 0.915 | 0.328 | 0.912 |
| 500 | 0.05 | MSE | 1.000 | 0.379 | 1.285 | 0.323 |
| 500 | 0.05 | \|Bias\| | 0.130 | 0.009 | 0.138 | 0.043 |
| 500 | 0.05 | Coverage | 0.767 | 0.925 | 0.615 | 0.938 |
| 1000 | 0.01 | MSE | 1.000 | 0.125 | 0.657 | 0.120 |
| 1000 | 0.01 | \|Bias\| | 0.300 | 0.009 | 0.24 | 0.080 |
| 1000 | 0.01 | Coverage | 0.058 | 0.875 | 0.13 | 0.858 |
| 1000 | 0.02 | MSE | 1.000 | 0.133 | 0.622 | 0.121 |
| 1000 | 0.02 | \|Bias\| | 0.220 | 0.003 | 0.169 | 0.048 |
| 1000 | 0.02 | Coverage | 0.240 | 0.922 | 0.31 | 0.915 |
| 1000 | 0.05 | MSE | 1.000 | 0.225 | 0.62 | 0.190 |
| 1000 | 0.05 | \|Bias\| | 0.118 | 0.008 | 0.087 | 0.021 |
| 1000 | 0.05 | Coverage | 0.598 | 0.940 | 0.642 | 0.948 |

*Table A 3 Monte Carlo performance of 2SLS, κ-weighting (normalized), PStrata, and Prince BART in Design C. Rows correspond to combinations of sample size $n$, overlap index $\delta$, and performance measure (relative MSE, absolute bias, and coverage). Columns correspond to the three estimators. Relative MSE is normalized so that 2SLS has value 1 in each $(n, \delta)$ scenario. Coverage refers to nominal 95% confidence intervals for 2SLS and κ-weighting and 95% posterior credible intervals for Prince BART. Results are based on 400 Monte Carlo replications.*

| n | delta | measure | 2SLS | $\kappa$-weighting (normalized) | PStrata | Prince BART |
|---|---|---|---|---|---|---|
| 500 | 0.01 | MSE | 1.000 | 0.106 | 0.742 | 0.116 |
| 500 | 0.01 | \|Bias\| | 0.570 | 0.021 | 0.505 | 0.173 |
| 500 | 0.01 | Coverage | 0.015 | 0.865 | 0.000 | 0.785 |
| 500 | 0.02 | MSE | 1.000 | 0.086 | 0.953 | 0.101 |
| 500 | 0.02 | \|Bias\| | 0.434 | 0.015 | 0.436 | 0.112 |
| 500 | 0.02 | Coverage | 0.080 | 0.907 | 0.010 | 0.853 |
| 500 | 0.05 | MSE | 1.000 | 0.116 | 1.046 | 0.107 |
| 500 | 0.05 | \|Bias\| | 0.264 | 0.000 | 0.276 | 0.055 |
| 500 | 0.05 | Coverage | 0.338 | 0.948 | 0.200 | 0.950 |
| 1000 | 0.01 | MSE | 1.000 | 0.043 | 0.713 | 0.049 |
| 1000 | 0.01 | \|Bias\| | 0.563 | 0.013 | 0.480 | 0.104 |
| 1000 | 0.01 | Coverage | 0.000 | 0.880 | 0.000 | 0.823 |
| 1000 | 0.02 | MSE | 1.000 | 0.038 | 0.688 | 0.043 |
| 1000 | 0.02 | \|Bias\| | 0.431 | 0.002 | 0.357 | 0.066 |
| 1000 | 0.02 | Coverage | 0.000 | 0.932 | 0.002 | 0.910 |
| 1000 | 0.05 | MSE | 1.000 | 0.051 | 0.523 | 0.060 |
| 1000 | 0.05 | \|Bias\| | 0.258 | 0.005 | 0.185 | 0.036 |
| 1000 | 0.05 | Coverage | 0.065 | 0.970 | 0.130 | 0.910 |

*Table A 4 Monte Carlo performance of 2SLS, κ-weighting (normalized), PStrata, and Prince BART in Design D. Rows correspond to combinations of sample size $n$, overlap index $\delta$, and performance measure (relative MSE, absolute bias, and coverage). Columns correspond to the three estimators. Relative MSE is normalized so that 2SLS has value 1 in each $(n, \delta)$ scenario. Coverage refers to nominal 95% confidence intervals for 2SLS and κ-weighting and 95% posterior credible intervals for Prince BART. Results are based on 400 Monte Carlo replications.*

## Appendix VII: Additional simulations and comparisons for the scaled Bayesian bootstrap

We implement a simulation and comparisons to check the performance of the scaled Bayesian bootstrap in the context of the DHS complex sampling design. Specifically, we compare the performance of the proposed scaled BB approach with a conventional design-based approach, which is the standard to conduct inference accounting for DHS complex sampling (Little, 2014). The covariate age is used as a hypothetical outcome of interest.

### *Simulation: DHS sample as a population*

For this simulation, we take the empirical distribution of X in the DHS Nigeria 2018 sample, say $F^*(x)$, as if it were the distribution in the population exactly. We take repeated samples from this population using a stratified two-stage cluster procedure as the one used to obtain the DHS sample, albeit with replacement.

Specifically, for each of the 72 strata, we take a sample of PSUs (the same size as in the DHS) with probability proportional to size (PPS), using the inverse of the average DHS sampling weight as a measure of size. For each selected PSU, a sample of respondents is selected by simple random samplings (SRS), again with the size corresponding to DHS sample (ranging from 26 to 40, with a median of 28). In both stages, sampling is done with replacement assuming a small sample fraction (1%).[11]

[11] The DHS second stage can involve a larger sample fraction. This information, however, is not publicly released and cannot be exploited in estimation.

For each replication we estimate the average age, denoted by $a$, in the hypothetical population, i.e., $\mu_a = \int a \, dF^*(a)$. We estimate $\mu_a$ using the proposed BB procedure (based on 1,000 bootstrap samples) and compare the results with those from a frequentist design-based estimator (implemented with the package survey, Lumley, 2020) with respect to bias, coverage, standard deviation and root mean square error.  For reference, we also include a "naïve" frequentist approach assuming SRS.

Table A 5 show the results based on 1,000 replications. The BB procedure generates estimates that perform very similar to those generated with the standard frequentist procedure. Reducing the number of clusters or strata did not seem to alter this conclusion (Table A 6 and Table A 7)

| | Bias | Coverage of 95% confidence or credible intervals | Std. deviation | Root mean square error |
|---|---|---|---|---|
| Naïve (i.e., assuming SRS) | .093 | .499 | .048 | .113 |
| Standard frequentist | -.003 | .954 | .065 | .064 |
| BB | -.003 | .968 | .073 | .064 |

*Table A 5 Performance of naïve (assuming SRS), standard design-based frequentist, and BB approaches for estimating the mean age in a population that is given by the DHS sample. Estimates are based on 1,000 replicated data sets using a stratified two-stage cluster sample with replacement from the true population.*

| | Bias | Coverage 95 %CIs | SD | RMSE |
|---|---|---|---|---|
| Naïve (i.e., assuming SRS) | .095 | .657 | .068 | .133 |
| Standard frequentist | -.001 | .951 | .093 | .093 |
| BB | -.001 | .97 | .104 | .093 |

*Table A 6 Performance of naïve (assuming SRS), standard design-based frequentist, and BB approaches for estimating the mean age in a population that is given by the DHS sample* ***with half the PSUs selected randomly****. Estimates are based on 1,000 replicated data sets using a stratified two-stage cluster sample with replacement from the true population.*

| | Bias | Coverage 95 %CIs | SD | RMSE |
|---|---|---|---|---|
| Naïve (i.e., assuming SRS) | .105 | .647 | .075 | .149 |
| Standard frequentist | .002 | .942 | .103 | .107 |
| BB | .002 | .963 | .116 | .107 |

*Table A 7 Performance of naïve (assuming SRS), standard design-based frequentist, and BB approaches for estimating the mean age in a population that is given by the DHS sample* ***with PSUs from urban strata only****. Estimates are based on 1,000 replicated data sets using a stratified two-stage cluster sample with replacement from the true population.*

***Comparison: Estimation of population outcomes using the DHS sample***

As an additional check, we compare estimation of simple quantities using DHS Nigeria 2018 using the BB procedure, a frequentist estimator assuming SRS (naïve) and a frequentist design-based estimator as in the simulation. Specifically, we focus on the average age, $\mu_a = \int a \, dF(a)$, where $F(a)$ is the distribution of age in the population, which is unknown. Unlike the case with simulation, in this case there is not a fixed value known in advance and thus we only present the different estimates.

Table A 8 shows the estimated quantities using different procedures. The cluster BB produces estimates that are practically identical to those obtained by the standard frequentist designed-based approach. Table A 9 include similar result when targeting a proportion.

| | Estimate | Std. deviation | Lower bound | Upper bound |
|---|---|---|---|---|
| Naïve (assuming SRS) | 32.00 | .092 | 31.82 | 32.18 |
| Standard frequentist | 31.59 | .133 | 31.33 | 31.85 |
| BB | 31.59 | .153 | 31.29 | 31.94 |

*Table A 8 Estimated mean age using naïve (assuming SRS), standard design-based frequentist, and BB approaches based on the actual DHS sample. Lower and upper bounds refer to bounds of 95% confidence or credible intervals.*

| | | | 95%CI | |
|---|---|---|---|---|
| | Est | SD | LW | UP |
| Naïve (assuming SRS) | .112 | .003 | .107 | .118 |
| Standard frequentist | .118 | .004 | .109 | .126 |
| BB | .118 | .005 | .108 | .127 |

*Table A 9 Estimated proportion of women 15 to 19 years old using naïve (assuming SRS), standard design-based frequentist, and BB approaches based on the actual DHS sample.*

**Appendix VIII: Additional descriptive characteristics**

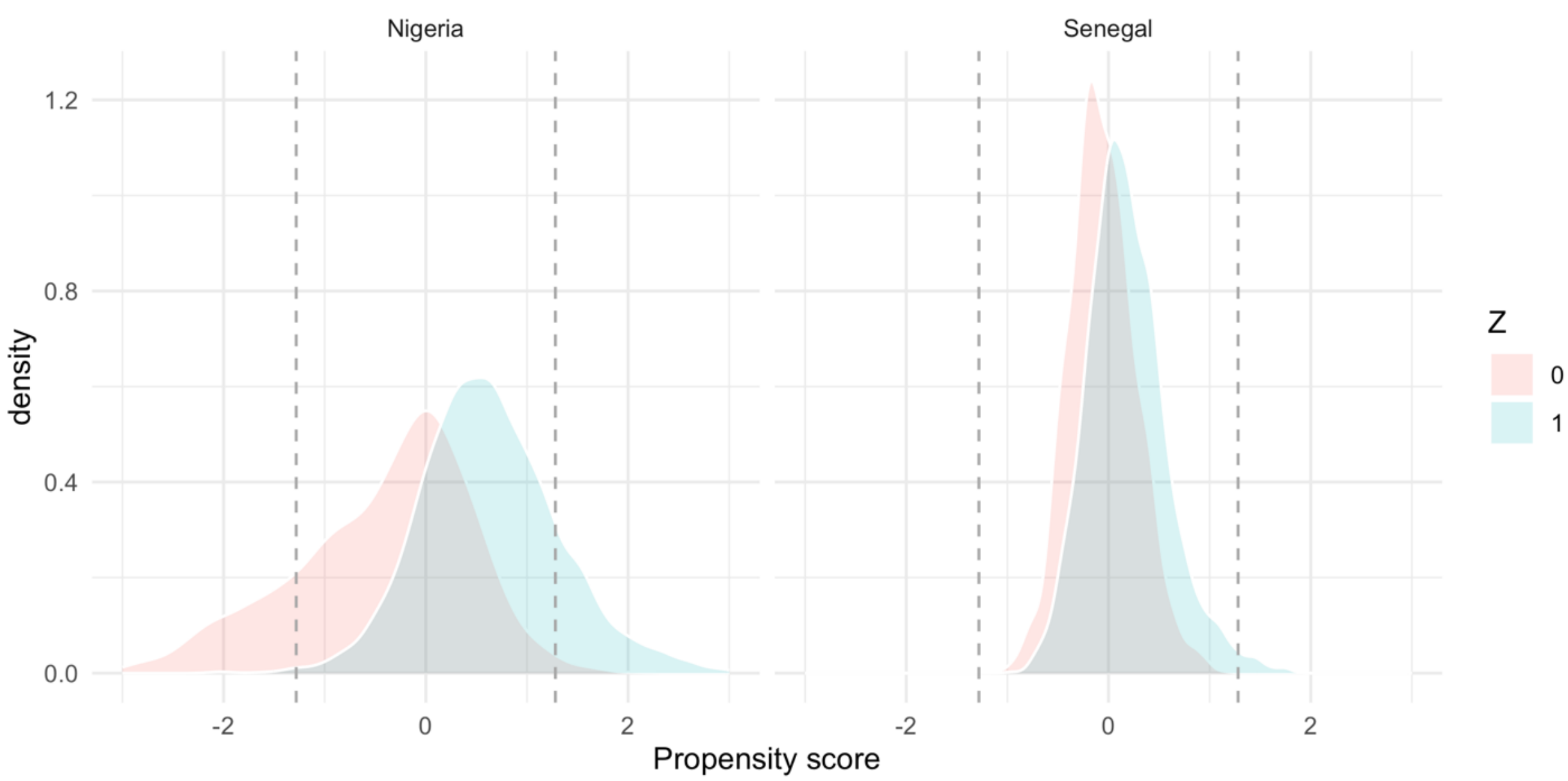

*Figure A 2 Distribution of the estimated propensity score (in probit scale) by assignment in Nigeria and Senegal. Dashed vertical lines correspond to 10% and 90% probability of assignment to treatment (Z=1). In the case of Nigeria, 17.3% of the sample has more extreme probabilities, in the case of Senegal, only .6%.*

| | Nigeria | | | Senegal | | |
|---|---|---|---|---|---|---|
| Baseline covariate | Mean Z=0 | Mean Z=1 | SMD | Mean Z=0 | Mean Z=1 | SMD |
| Age | 26.593 | 26.936 | 0.037 | 25.667 | 25.837 | 0.019 |
| att_autonomous_use | 0.174 | 0.156 | -0.047 | 0.167 | 0.206 | 0.101 |
| att_safety | 0.409 | 0.388 | -0.045 | 0.445 | 0.366 | -0.161 |
| edu_higher | 0.130 | 0.184 | 0.149 | 0.014 | 0.040 | 0.157 |
| edu_none | 0.161 | 0.110 | -0.149 | 0.404 | 0.311 | -0.194 |
| edu_primary | 0.167 | 0.146 | -0.059 | 0.285 | 0.339 | 0.118 |
| edu_secondary_comp | 0.360 | 0.429 | 0.143 | 0.079 | 0.087 | 0.028 |
| edu_secondary_incomp | 0.177 | 0.122 | -0.153 | 0.218 | 0.223 | 0.013 |
| FP_radio | 0.485 | 0.374 | -0.225 | 0.187 | 0.174 | -0.034 |
| FP_TV | 0.301 | 0.290 | -0.024 | 0.296 | 0.403 | 0.226 |
| had_sex | 0.642 | 0.624 | -0.037 | 0.522 | 0.459 | -0.125 |
| has_money | 0.539 | 0.525 | -0.029 | 0.505 | 0.429 | -0.153 |
| knwl_contraception | 0.923 | 0.867 | -0.185 | 0.962 | 0.975 | 0.075 |
| marstat_divorced | 0.036 | 0.029 | -0.036 | 0.052 | 0.058 | 0.025 |
| marstat_married | 0.582 | 0.550 | -0.065 | 0.456 | 0.371 | -0.175 |
| marstat_never_married | 0.378 | 0.411 | 0.066 | 0.492 | 0.572 | 0.161 |
| paid_cash | 0.431 | 0.419 | -0.024 | 0.377 | 0.364 | -0.027 |
| parity | 2.395 | 1.968 | -0.159 | 1.446 | 1.158 | -0.131 |
| religion_Christian | 0.342 | 0.408 | 0.136 | 0.025 | 0.059 | 0.169 |
| religion_Muslim | 0.644 | 0.584 | -0.125 | 0.973 | 0.938 | -0.173 |
| self_employed | 0.384 | 0.326 | -0.122 | 0.262 | 0.219 | -0.101 |
| selfeff_obtain | 0.545 | 0.602 | 0.115 | 0.771 | 0.756 | -0.034 |
| teen_birth | 0.202 | 0.117 | -0.233 | 0.123 | 0.083 | -0.132 |
| want_no_birth | 0.535 | 0.505 | -0.061 | 0.672 | 0.695 | 0.051 |
| wealth_middle | 0.211 | 0.192 | -0.046 | 0.233 | 0.248 | 0.035 |
| wealth_poorer | 0.216 | 0.197 | -0.047 | 0.202 | 0.216 | 0.032 |
| wealth_poorest | 0.229 | 0.173 | -0.141 | 0.203 | 0.159 | -0.116 |
| wealth_richer | 0.202 | 0.219 | 0.041 | 0.206 | 0.197 | -0.024 |
| wealth_richest | 0.142 | 0.219 | 0.201 | 0.156 | 0.182 | 0.069 |
| work_last_week | 0.361 | 0.402 | 0.083 | 0.348 | 0.315 | -0.071 |
| work_last_year | 0.463 | 0.455 | -0.017 | 0.440 | 0.437 | -0.006 |

*Table A 10 Baseline covariate means and standardized mean differences (SMD) between women assigned to FP program early roll out (Z=1) and women assigned to control (Z=0) in the MLE source sample, for Nigeria and Senegal.*

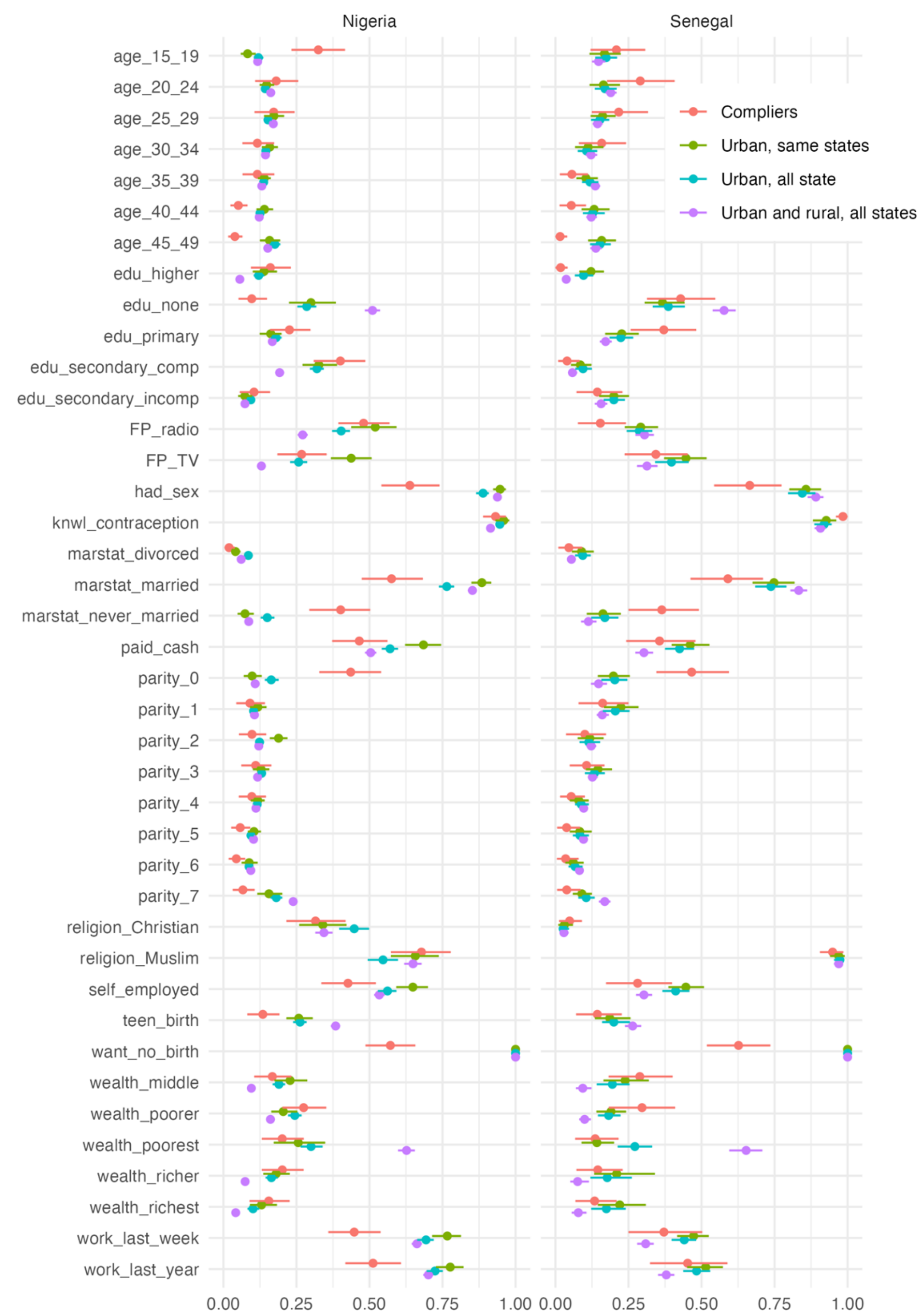

*Figure A 3 Descriptive characteristics for four populations: (1) compliers in the source sample, and (2) all women residing in urban areas in the same states/regions represented in the source sample, (3) all women residing in urban areas, and (4) all women in the country, for Nigeria and Senegal.*

| var | compliers in the source | urban areas in the same states | | urban areas, all states | | Urban and rural, all states | |
|---|---|---|---|---|---|---|---|
| | mean | mean | SMD | mean | SMD | mean | SMD |
| age | 26.069 | 32.602 | 0.717 | 32.322 | 0.661 | 31.589 | 0.591 |
| Christian | 0.319 | 0.338 | 0.041 | 0.447 | 0.266 | 0.343 | 0.052 |
| edu_higher | 0.160 | 0.138 | -0.064 | 0.121 | -0.114 | 0.056 | -0.340 |
| edu_junioHS | 0.109 | 0.073 | -0.124 | 0.094 | -0.050 | 0.074 | -0.122 |
| edu_primary | 0.222 | 0.161 | -0.154 | 0.180 | -0.104 | 0.167 | -0.138 |
| edu_seniorHS | 0.400 | 0.326 | -0.154 | 0.320 | -0.167 | 0.192 | -0.466 |
| FP_radio | 0.478 | 0.520 | 0.086 | 0.403 | -0.151 | 0.271 | -0.436 |
| FP_TV | 0.272 | 0.437 | 0.351 | 0.257 | -0.033 | 0.130 | -0.360 |
| had_sex | 0.637 | 0.947 | 0.826 | 0.889 | 0.618 | 0.938 | 0.789 |
| in_union | 0.575 | 0.885 | 0.746 | 0.764 | 0.411 | 0.852 | 0.645 |
| knwl_contraception | 0.926 | 0.959 | 0.139 | 0.946 | 0.081 | 0.914 | -0.044 |
| Muslim | 0.673 | 0.659 | -0.030 | 0.547 | -0.260 | 0.650 | -0.049 |
| never_married | 0.397 | 0.074 | -0.824 | 0.150 | -0.576 | 0.087 | -0.777 |
| no_edu | 0.101 | 0.302 | 0.517 | 0.285 | 0.481 | 0.511 | 0.993 |
| paid_cash | 0.466 | 0.684 | 0.452 | 0.569 | 0.207 | 0.504 | 0.075 |
| parity | 2.125 | 3.738 | 0.606 | 3.777 | 0.591 | 4.313 | 0.766 |
| self_employed | 0.422 | 0.649 | 0.467 | 0.561 | 0.280 | 0.533 | 0.224 |
| separated | 0.022 | 0.041 | 0.109 | 0.086 | 0.284 | 0.061 | 0.195 |
| teen_birth | 0.138 | 0.259 | 0.307 | 0.262 | 0.315 | 0.384 | 0.585 |
| want_no_birth | 0.567 | 1.000 | 1.235 | 1.000 | 1.235 | 1.000 | 1.235 |
| wealth_1 | 0.202 | 0.256 | 0.129 | 0.300 | 0.227 | 0.627 | 0.955 |
| wealth_2 | 0.269 | 0.205 | -0.151 | 0.244 | -0.056 | 0.161 | -0.265 |
| wealth_3 | 0.169 | 0.229 | 0.148 | 0.190 | 0.054 | 0.096 | -0.219 |
| wealth_4 | 0.202 | 0.180 | -0.055 | 0.164 | -0.098 | 0.074 | -0.377 |
| wealth_5 | 0.157 | 0.130 | -0.078 | 0.101 | -0.168 | 0.042 | -0.391 |
| work_last_week | 0.444 | 0.766 | 0.698 | 0.693 | 0.518 | 0.662 | 0.450 |
| work_last_year | 0.512 | 0.776 | 0.576 | 0.724 | 0.447 | 0.701 | 0.396 |

*Table A 11 Baseline covariate means and standardized mean differences (SMD) comparing compliers in the MLE source sample with each of the DHS-based target populations Nigeria.*

| | compliers in the source | urban areas in the same states | | urban areas, all states | | Urban and rural, all states | |
|---|---|---|---|---|---|---|---|
| | mean | mean | SMD | mean | mean | mean | SMD |
| age | 26.069 | 32.602 | 0.717 | 32.322 | 0.661 | 31.589 | 0.591 |
| Christian | 0.319 | 0.338 | 0.041 | 0.447 | 0.266 | 0.343 | 0.052 |
| edu_higher | 0.160 | 0.138 | -0.064 | 0.121 | -0.114 | 0.056 | -0.340 |
| edu_junioHS | 0.109 | 0.073 | -0.124 | 0.094 | -0.050 | 0.074 | -0.122 |
| edu_primary | 0.222 | 0.161 | -0.154 | 0.180 | -0.104 | 0.167 | -0.138 |
| edu_seniorHS | 0.400 | 0.326 | -0.154 | 0.320 | -0.167 | 0.192 | -0.466 |
| education | 3.269 | 2.535 | -0.430 | 2.526 | -0.444 | 1.606 | -1.001 |
| FP_radio | 0.478 | 0.520 | 0.086 | 0.403 | -0.151 | 0.271 | -0.436 |
| FP_TV | 0.272 | 0.437 | 0.351 | 0.257 | -0.033 | 0.130 | -0.360 |
| had_sex | 0.637 | 0.947 | 0.826 | 0.889 | 0.618 | 0.938 | 0.789 |
| in_union | 0.575 | 0.885 | 0.746 | 0.764 | 0.411 | 0.852 | 0.645 |
| knwl_contraception | 0.926 | 0.959 | 0.139 | 0.946 | 0.081 | 0.914 | -0.044 |
| Muslim | 0.673 | 0.659 | -0.030 | 0.547 | -0.260 | 0.650 | -0.049 |
| never_married | 0.397 | 0.074 | -0.824 | 0.150 | -0.576 | 0.087 | -0.777 |
| no_edu | 0.101 | 0.302 | 0.517 | 0.285 | 0.481 | 0.511 | 0.993 |
| paid_cash | 0.466 | 0.684 | 0.452 | 0.569 | 0.207 | 0.504 | 0.075 |
| parity | 2.125 | 3.738 | 0.606 | 3.777 | 0.591 | 4.313 | 0.766 |
| self_employed | 0.422 | 0.649 | 0.467 | 0.561 | 0.280 | 0.533 | 0.224 |
| separated | 0.022 | 0.041 | 0.109 | 0.086 | 0.284 | 0.061 | 0.195 |
| teen_birth | 0.138 | 0.259 | 0.307 | 0.262 | 0.315 | 0.384 | 0.585 |
| want_no_birth | 0.567 | 1.000 | 1.235 | 1.000 | 1.235 | 1.000 | 1.235 |
| wealth_1 | 0.202 | 0.256 | 0.129 | 0.300 | 0.227 | 0.627 | 0.955 |
| wealth_2 | 0.269 | 0.205 | -0.151 | 0.244 | -0.056 | 0.161 | -0.265 |
| wealth_3 | 0.169 | 0.229 | 0.148 | 0.190 | 0.054 | 0.096 | -0.219 |
| wealth_4 | 0.202 | 0.180 | -0.055 | 0.164 | -0.098 | 0.074 | -0.377 |
| wealth_5 | 0.157 | 0.130 | -0.078 | 0.101 | -0.168 | 0.042 | -0.391 |
| work_last_week | 0.444 | 0.766 | 0.698 | 0.693 | 0.518 | 0.662 | 0.450 |
| work_last_year | 0.512 | 0.776 | 0.576 | 0.724 | 0.447 | 0.701 | 0.396 |

*Table A 12 Baseline covariate means and standardized mean differences (SMD) comparing compliers in the MLE source sample with each of the DHS-based target populations Senegal.*

**Appendix IX: Effect heterogeneity**

| segment | Estimate (standard deviation) | 90%CI |
|---|---|---|
| Overall (MATE) | 0.366 (0.133) | 0.150, 0.585 |
| never married, medium-highest wealth, did not work last year | -0.020 (0.222) | -0.394, 0.330 |
| never married, lower and lowest wealth, did not work last year | 0.192 (0.227) | -0.196, 0.541 |
| never married, medium-highest wealth, work last year | 0.254 (0.211) | -0.108, 0.587 |
| married or separated, did not work last year, medium-highest wealth | 0.334 (0.205) | -0.010, 0.667 |
| never married, lower and lowest wealth, work last year | 0.418 (0.200) | 0.067, 0.720 |
| married or separated, did not work last year, lower and lowest wealth | 0.466 (0.189) | 0.142, 0.768 |
| married or separated, work last year, some secondary education or higher | 0.532 (0.158) | 0.274, 0.799 |
| married or separated, work last year, primary education or less | 0.647 (0.133) | 0.418, 0.854 |

*Table A 13 Effect of contraceptive use on employment as a function of selected combination of covariates,* $MCATE_c$*, in Nigeria*

| segment | Estimate (standard deviation) | 90%CI |
|---|---|---|
| Overall (MATE) | 0.473 (0.155) | 0.202, 0.706 |
| not sexually active, did not work last year, less than 20 years old | 0.267 (0.230) | -0.112, 0.651 |
| not sexually active, did not work last year, 20 years or more | 0.353 (0.217) | -0.023, 0.690 |
| not sexually active, work last year, at least some secondary education | 0.380 (0.226) | -0.021, 0.720 |
| not sexually active, work last year, primary education or less | 0.456 (0.207) | 0.086, 0.768 |
| sexually active, 30 or more years old, did not work last year | 0.472 (0.219) | 0.081, 0.799 |
| sexually active, less than 30 years old, at least some secondary education | 0.493 (0.207) | 0.119, 0.796 |
| Sexually active, less than 30 years old, at least some secondary education | 0.525 (0.206) | 0.173, 0.839 |
| Sexually active, less than 30 years old, primary education or less | 0.575 (0.170) | 0.268, 0.820 |

*Table A 14 Effect of contraceptive use on employment as a function of selected combination of covariates,* $MCATE_c$*, in Senegal*

**Appendix X: Sensitivity to conservative single imputation of FP attitudes**

In the main analysis, the four FP attitude covariates that are observed in the MLE source data but missing in the DHS target data are treated as latent and imputed for DHS women using BART models fit in the source sample (see Methods, "Missing covariates"). This BART-based multiple imputation propagates both binomial and model uncertainty about these attitudes into the PATE posterior.

As a sensitivity analysis, we consider the more conservative single-imputation strategy used in our original submission. For each of the four attitude variables, we previously examined how the estimated $MATE_C$ in the source sample changed when that variable was set to 0 or 1 for all women (Figure A 4- Figure A 5). Based on these "partial effects," we selected the combination of attitude values associated with the smallest estimated complier effect and assigned that pattern to all DHS women. Table A15 reports the resulting PATE estimates for Nigeria and Senegal.

Compared with the BART-based imputation in the main analysis (Table A 15), the conservative single-imputation scenario yields slightly smaller generalized effects and somewhat wider posterior uncertainty. For Nigeria, PATE estimates decrease by about 0.03–0.04 across target populations (e.g., from roughly 0.50–0.53 to 0.46–0.49), and standard deviations increase by about 0.02–0.03. For Senegal, PATE estimates decrease from about 0.48–0.49 to about 0.45, with similar modest increases in uncertainty. In all cases, the estimated effects remain positive and substantial, and the qualitative conclusions are unchanged, suggesting that our findings are not driven by the particular treatment of these four attitude covariates.

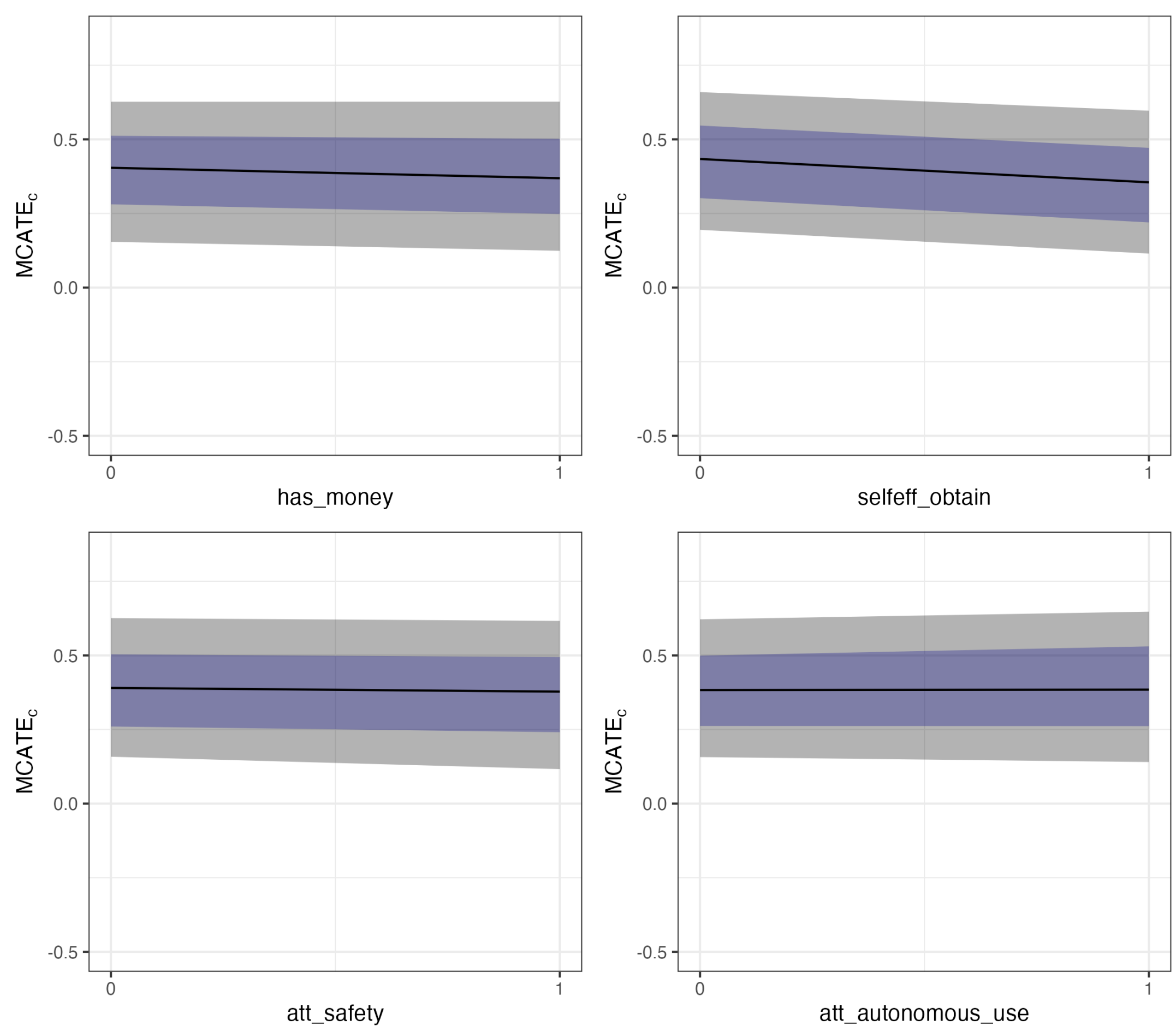

*Figure A 4 Effect of contraceptive use on employment in Nigeria's source sample as a function of a change in a single covariate. The black line is the point estimate, while blue and gray bands represent 60 and 90% credible intervals.*

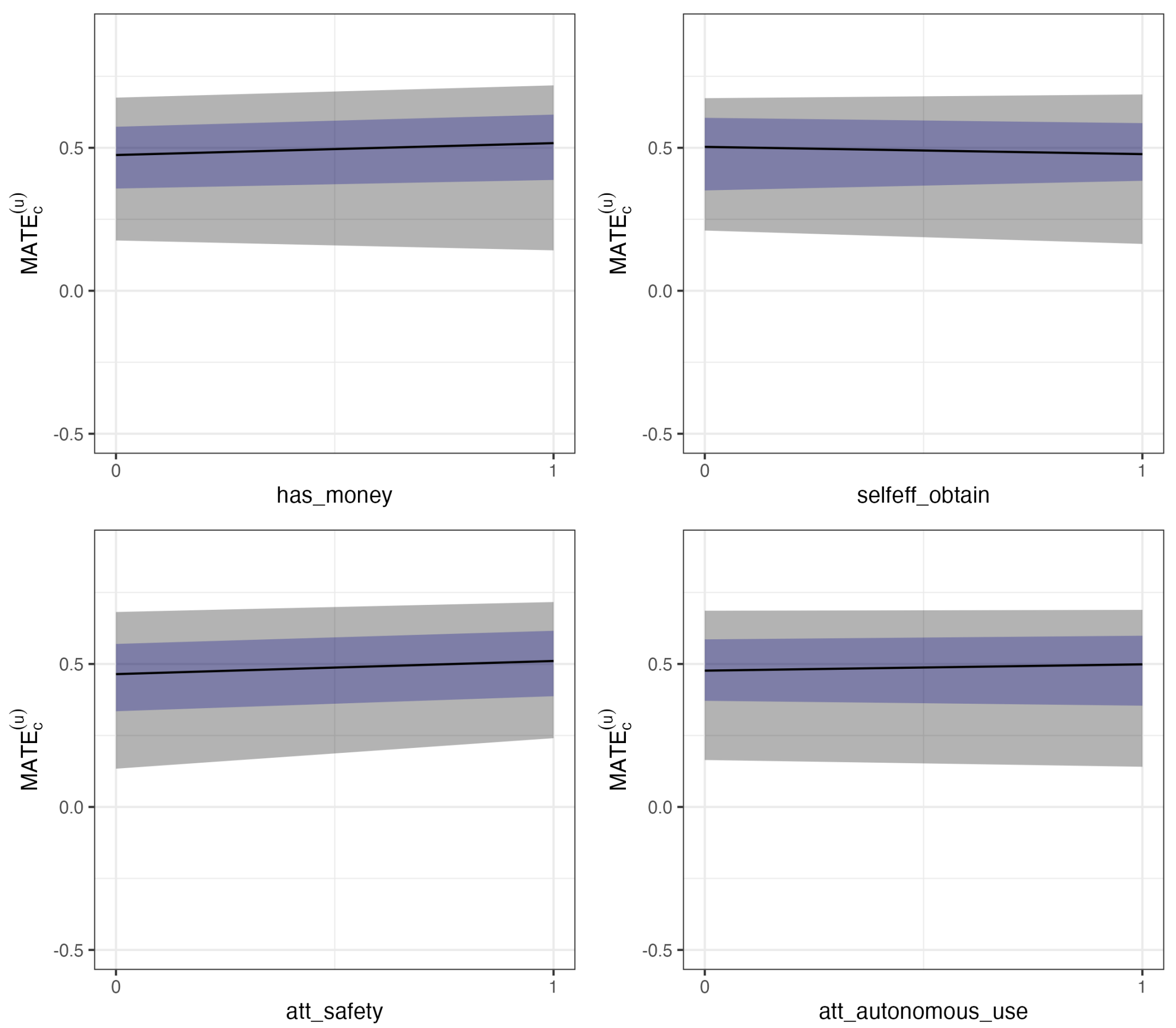

*Figure A 5 Effect of contraceptive use on employment in Senegal's source sample as a function of a change in a single covariate. The black line is the point estimate, while blue and gray bands represent 60 and 90% credible intervals.*

| | Nigeria | Senegal |
|---|---|---|
| $MATE_C$ (Prince BART) | 0.366 (0.133) | 0.473 (0.155) |
| PATE | | |
| Urban, same states | 0.478 (0.174) | 0.454 (0.200) |
| Urban, all states | 0.456 (0.173) | 0.449 (0.201) |
| Urban and rural, all states | 0.485 (0.176) | 0.450 (0.214) |

*Table A 15 Effect of adopting modern contraception on employment: average effect among compliers in the source study ($MATE_C$, Prince BART) and generalized effects (PATE) for different target populations under the conservative single-imputation scenario for FP attitude covariates.*

**Appendix XI: Sensitivity to violation of conditional transportability (assumption 5) – a parametric approach**

The sensitivity analysis in the body of the text avoids parametric assumption about the unobserved effect modifier. Because the additional flexibility, point identification is lost, and the interpretation is less straightforward. Here we present an alternative sensitivity analysis in which we pose a relatively simple parametric model for the unobserved effect modifier. We assume that $U_i$ is a binary predictor, unrelated with $X_i$, that affects the CATE's additively (except near boundaries, to ensure that CATEs are kept within their range). The model set up can be summarize as follows,

$$\ddot{PATE}^{\xi} = \mathbb{E}_{X,U|T=1}\big(\mathrm{CATE}'(\mathrm{X_i},\mathrm{U_i})\big),$$

$$\mathrm{CATE}'(\mathrm{x},\mathrm{u}) = h(\mathbb{E}(\mathrm{Y_i}|\mathrm{W_i}=1,\mathrm{X_i}) - \mathbb{E}(\mathrm{Y_i}|\mathrm{W_i}=0,\mathrm{X_i}) + U_i\cdot\kappa)$$

$$U_i|T_i = 1 \sim Bern(\xi),$$

*(A 4)*

where $h(x) = \max(\min(x,1), -1)$ ensures that the modified CATEs remain within plausible range. We fix $\kappa$, and vary the prevalence of the confounder in the population, $\xi$. The value of $\kappa$ is set to 0.37 in Nigeria and .18 in Senegal, the maximum difference between the effect in one of the identified segments from the average.

Figure 10 presents the relation between the estimated PATE and population prevalence of a strong of confounder which is absent in the compliers sample. The confounder is as strong as the strongest observed effect modifier. The findings suggest that a strong confounder should be present in over half of Nigerian or Senegalese women, for the 95% CI of the PATE to include zero.

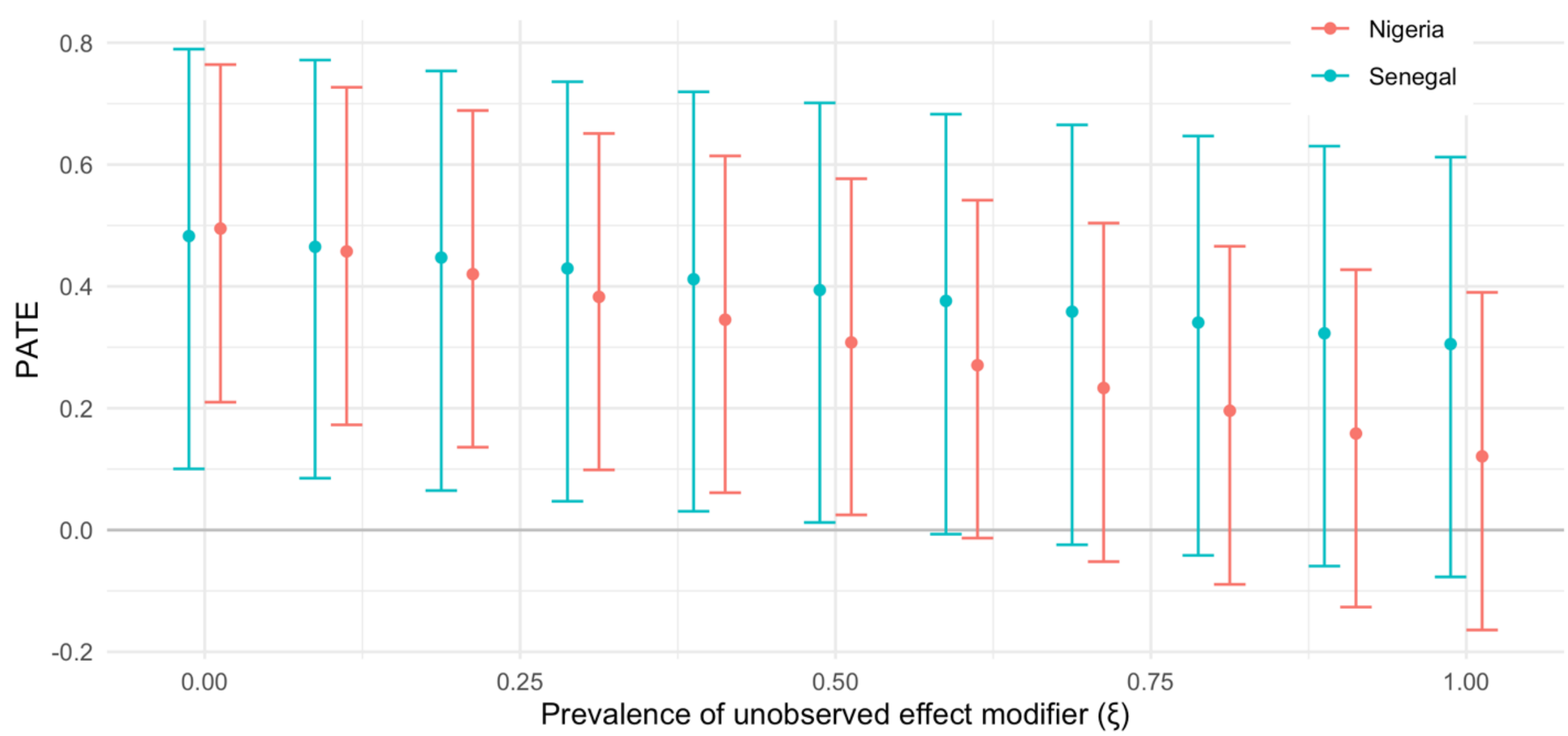

*Figure 10 Sensitivity of the PATE to population prevalence of a strong of effect modifier.*